\documentclass[fleqn,usenatbib]{mnras}
\usepackage[T1]{fontenc}

\DeclareRobustCommand{\VAN}[3]{#2}
\let\VANthebibliography\thebibliography
\def\thebibliography{\DeclareRobustCommand{\VAN}[3]{##3}\VANthebibliography}

\usepackage{graphicx}	% Including figure files
\usepackage{amsmath}	% Advanced maths commands
\usepackage{amssymb}	% Extra maths symbols
\usepackage{hyperref}
\usepackage{color}
\usepackage{multirow}
\usepackage{newtxtext,newtxmath}
\graphicspath{{./}{figures/}}

\newcommand\bpmrp{G_{{\rm BP}}-G_{{\rm RP}}}
\newcommand\grp{G-G_{{\rm RP}}}
\newcommand\vt{v_{t}}
\newcommand\masyr{\rm mas\,yr^{-1}}
\newcommand\kms{\rm km\,s^{-1}}
\DeclareUnicodeCharacter{2212}{-}
\title[Stars in the Local Galactic Thick Disk and Halo]{Stars in the Local Galactic Thick Disk and Halo in Gaia EDR3: A catalogue of Half a Million Local Main-sequence Stars with Photometric Metallicities}

\author[Kim \& L{\'e}pine]{
Bokyoung Kim$^{1}$\thanks{E-mail: bokyoung.kim@ed.ac.uk}
and Sebastien L{\'e}pine$^{2}$
\\
$^{1}$Institute for Astronomy, University of Edinburgh, Royal Observatory, Blackford Hill, Edinburgh EH9 3HJ, UK\\
$^{2}$Department of Physics and Astronomy, Georgia State University, 25 Park Place, Atlanta, GA 30303, USA\\
}

\date{Accepted 2021 December 13. Received 2021 December 13; in original form 2021 September 11}

\pubyear{2021}

\begin{document}
\label{firstpage}
\pagerange{\pageref{firstpage}--\pageref{lastpage}}
\maketitle

% Abstract of the paper
\begin{abstract}
We present a catalogue of $551,214$ main-sequence stars in the local ($d < 2$ kpc) Galactic thick disk and halo, based on a search of stars with large proper motions ($\mu_{\rm tot} > 40.0~\masyr$) in the Gaia Early Data Release 3. We derive photometric metallicity calibrated from the colour-luminosity-metallicity distribution of $20,047$ stars with spectroscopic metallicities, collected from various spectroscopic surveys, including SDSS SEGUE/APOGEE, GALAH DR3, and LAMOST DR6. We combine these results to construct an empirical colour-magnitude-metallicity grid, which can be used to estimate photometric metallicities for low-mass metal-poor stars of K and M subtypes from their absolute $G$ magnitude and colour values. We find that low-mass, high-velocity stars in our catalogue share similar kinematics as reported in recent studies of more luminous Galactic halo stars. The pseudo-kinematic analysis of our sample recovers the main local halo structures, including the Gaia-Enceladus Stream and the Helmi stream; aside from these the local halo stars appear to show a remarkably smooth distribution in velocity space. Since the future Gaia data release will provide radial velocity measurements for only a small number of our sample, our catalogue provides targets of high interest for the future spectroscopic observation programs like SDSS-V, DESI MW survey, WEAVE, and/or 4MOST.
\end{abstract}

% Select between one and six entries from the list of approved keywords.
% Don't make up new ones.

%%%%%%%%%%%%%%%%%%%%%%%%%%%%%%%%%%%%%%%%%%%%%%%%%%

%% Start the main body of the article. If no sections in the 
%% research note leave the \section call blank to make the title.
\section{Introduction}\label{sec:intro}
Thanks to the Gaia mission \citep{gaia:16, gaia:18a, gaia:20}, our understanding of the Milky Way has deepened, with an accumulation of new information and detailed data that inform the formation and evolutionary history of the Galaxy. In itself, the second Gaia data release \citep[DR2,][]{gaia:18a}, now provides a vast dataset to survey and map out the stellar populations of the Galactic disk and halo. However, Gaia data sets become even more powerful when they are combined with other existing large spectroscopic datasets that provide complementary information on radial velocity and elemental abundance. For example, recent studies \citep{belokurov:18, helmi:18, myeong:18a, myeong:18b, koppelman:18} based on Gaia DR2 and the Sloan Digital Sky Survey \citep[SDSS,][]{blanton:17, yanny:09, majewski:17}  have reported a slightly retrograde, large ellipsoidal structure in velocity space that appears to have been created by an accretion event involving a massive dwarf galaxy, known as the Gaia-Enceladus-Sausage or Gaia-Enceladus\footnote{Hereafter, we use the term of "Gaia-Enceladus" to describe the structure, if necessary.}. \citet{haywood:18} investigated an apparent doubling of the main-sequence in the colour-magnitude diagram (CMD) of stars with high tangential velocities in Gaia DR2 and reported that the halo populations found in the solar vicinity are a combination of accreted stars and of the tail of the thick disk that was dynamically heated by past accretion events. \citet{naidu:20} identified seven known substructures tied to recently identified accretion events \citep{horta:21, koppelman:19b, myeong:18c, myeong:18a} and three new structures all within $50$ kpc from the Galactic centre, using a sample of giants at high Galactic latitude for which radial velocities and elemental abundances are obtained from the H3 spectroscopic survey \citep{conroy:19}.

Most results probing the spatial structure and kinematics of the Galactic halo have so far been obtained from the analysis of bright sources, mainly giants, which are the best targets for observations aiming to study the kinematics of moderately distant halo stars as they require less effort to obtain high signal-to-noise spectral data for measuring radial velocities and elemental abundances. Moreover, using intrinsically bright sources as tracers of the halo populations maximises the volume that one can survey, which can unveil the structure of the halo on the largest scales (tens of kiloparsecs) to probe the global formation and chemical enrichment history of the halo, and identify major evolutionary events. Bright sources, therefore, are the most favourable targets for studying the outer regions of the Galaxy.

However, the main constituents of the stellar halo are not evolved objects like red giants but main-sequence stars. Low-mass stars in particular (K and M dwarfs) make up the bulk of the baryonic halo, but they remain largely unexploited in studies of the Galactic halo due to their low intrinsic luminosity. Due to these technical difficulties, surveys of main-sequence stars merely extend out to a few kiloparsecs from the Sun, and thus only probe the {\em local} halo populations. Nevertheless, a merit of using low-mass main-sequence stars is that they are ``pristine" and have preserved the original chemical compositions of their birthplace in their atmosphere. As low-mass stars stay in the main-sequence stage for much longer than $10$ Gyr, those stars in the local halo effectively deliver accurate abundance information about the star-forming environment of the early Universe \citep{frebel:15, helmi:20}, and hence provide valuable records of the Solar vicinity in the early Galaxy. In addition, since low-mass stars are $>10^2$ times more common than red giants \citep{lepine:05a, lepine:05b, henry:06}, they can potentially reveal the signature of much smaller accretion or formation events, drawing a more complete picture of Galaxy evolution.

The latest Gaia Early Data Release 3 \citep[EDR3,][]{gaia:20} provides parallaxes and proper motions for more than $1.48$ billion stars in the Milky Way. The magnitude limit in the $G$ band is $\sim 21.5$, which makes it possible to identify and characterise millions of low-mass stars in the Solar neighbourhood and obtain their precise distances, proper motions, and colours. Although Gaia cannot provide full 6D astrometric solutions (i.e., position, proper motion, distance, and radial velocity) for a large fraction of the low-mass stars, which tend to be fainter than $G \sim 16$, Gaia EDR3 expands the amount of main-sequence stars that we can use to statistically investigate the kinematics of the local halo. With a survey completeness limit at $G \le 19$ \citep{fabricius:20}, we can identify even very low-mass stars (e.g. late-type M dwarfs) with absolute magnitude $M_{G}\sim 13$ up to a distance $d < 150$~pc. Gaia EDR3 reports an average improvement on the uncertainties for the positions and parallaxes by a factor of $\sim 0.8$ and for the proper motions by a factor of $0.5$ \citep{lindegren:21} with respect to Gaia DR2, which is thanks to the longer baseline from $33$ months of observations as compared to $22$-month data for Gaia DR2 \citep{lindegren:21, fabricius:20}. The significant improvement in Gaia astrometry benefits the identification of the faint halo population since kinematic selection of high-velocity stars is the primary method for rounding up halo stars.

The application of simple kinematic cuts for selecting halo main-sequence stars is not sufficient as the faster-moving stars from the Galactic thick disk population will typically contaminate the sample in significant numbers. Recent studies have reported that stars in the Solar neighbourhood with high tangential velocities ($\vt > 200$ km s$^{-1}$) primarily consist of a mixture of stars from the Gaia-Enceladus dwarf galaxy that merged with the Milky Way $9$ -- $11$ Gyr ago and stars from the old disk heated by the Gaia-Enceladus accretion event \citep{belokurov:18, helmi:18, koppelman:18, haywood:18}. Since both populations have different origins, their chemical make-up also differs significantly and they display different ranges of iron abundances [Fe/H], which can serve as a clue to their origin. True halo stars, either from the Gaia-Enceladus, or from other earlier accretion events, are typically the most metal-poor, and their chemical make up can serve to identify them as it affects their spectral distribution.

In the absence of spectroscopic data, an indirect approach to identify metal-poor stars is to use known relationships between metallicity, colour, and luminosity, which can be applied in a CMD. Empirically, metal-poor stars on the main sequence are both bluer and brighter in optical CMDs compared to metal-rich stars. As large spectroscopic surveys, like SDSS SEGUE \citep{yanny:09}, APOGEE \citep{majewski:17, ahumada:20}, GALAH \citep{desilva:15, buder:20}, and LAMOST \citep{cui:12}, have collected and provided a huge collection of stellar parameters in the last decade, including metallicity measurements, data for these stars can be used to calibrate colour-magnitude-metallicity relationships, from which {\em photometric metallicities} can be estimated empirically. There have been many studies reporting photometric metallicities derived using metallicity-sensitive photometry, such as SDSS $ugriz$ filter photometry \citep{ivezic:08, an:13} or SkyMapper $v$ filter on the sensitivity of the Ca II K absorptions connecting to $ugi$ photometry \citep{casagrande:19, onken:19, chiti:21}. While most of these studies have been done for higher-mass stars, empirical calibration of colour-metallicity-luminosity relationships for low-mass stars is still in progress, and depends on assembling spectroscopic data for large enough numbers of low-mass stars.

In this paper, we assemble a large catalogue of candidate halo main-sequence stars within $2$ kpc from the Sun from Gaia EDR3. A search of recent, large spectroscopic surveys of low-mass stars identifies $>17,000$ stars that have measured spectroscopic metallicities; we use these stars to calibrate a {\em metallicity grid} in the optical CMD, which we use to estimate photometric metallicities for all the stars in the catalogue. In \S\ref{sec:data}, we discuss the processes to identify candidates in the local Galactic halo from Gaia EDR3 and introduce additional datasets we collected to build the photometric metallicity grid. In \S\ref{sec:phot_feh}, we present how to estimate photometric metallicities of our sample and how to calibrate the grid based on common-proper-motion pairs in our catalogue. Detailed information and validation of our catalogue is provided in \S\ref{sec:catalogue}, which includes estimates of the local density of halo stars and their luminosity function. We summarise our results in \S\ref{sec:summary}.

\section{Data}\label{sec:data}
\subsection{A catalogue of High-proper-motion Stars in Gaia EDR3}

\begin{figure*}
\includegraphics[width=\textwidth]{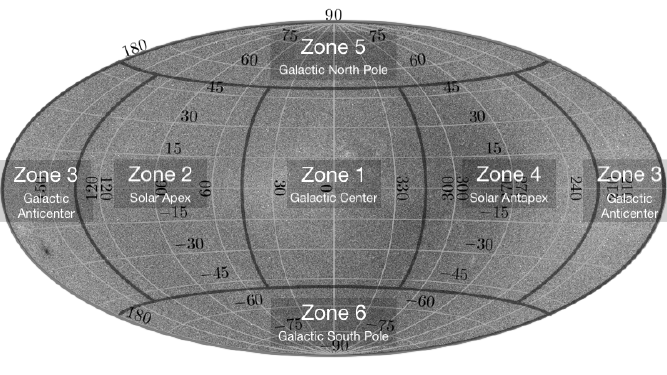}
\caption{Sky distribution of the full subset of $3.40$ million stars with precise Gaia parallaxes ($\sigma_{\pi}/\pi < 0.150$) and high-proper-motions ($\mu_{\mathrm {tot}} > 40.0$ mas yr$^{-1}$). We define six zones that correspond to the six cardinal directions in the Galactic coordinate system. Due to the ``asymmetric drift", halo stars have specific distributions of their transverse motions in each sector. \label{fig:fig1}}
\end{figure*}

Gaia EDR3 provides parallax, proper motion, and colour for $\sim 1.46$ billion stars, which is the largest collection of data on stars in the Galaxy. To select stars from the local halo populations, we first assemble a list of stars with precise Gaia parallaxes ($\sigma_{\pi}/\pi < 0.15$) and reported high proper motions ($\mu > 40.0$~mas yr$^{-1}$). A chance to catch halo stars increases, using this particular proper motion limit; for example, a tangential velocity of a star with $\mu = 40.0$ mas yr$^{-1}$ at the distance of $1$ kpc is $189.6$ km s$^{-1}$. We further introduce a set of conditions to extract a subset with consistent and reliable astrometric and photometric data. \citet{lindegren:18, lindegren:21} recommended checking uncertainties in the parallax measurements, uncertainties in fluxes of the Gaia BP and RP filters, the renormalised unit weight errors (RUWE), and $\bpmrp$~colour excess factors ($E$):
\begin{enumerate}
\item $\pi > 0.0$
\item $\sigma_{\pi}/\pi < 0.15$
\item $\mu_{\rm total} > 40.0$~mas yr$^{-1}$
\item $-3.0 \le \bpmrp < 6.0$
\item $3.0 \le G < 21.0$
\item $\sigma(F_{{\rm BP}})/F_{{\rm BP}} < 0.10$
\item $\sigma(F_{\rm {RP}})/F_{\rm {RP}} < 0.10$
\item ${\rm RUWE} < 1.4$
\item $1.0 + 0.015 (G_{{\rm BP}}-G_{{\rm RP}})^2 < E < 1.3 + 0.06(G_{{\rm BP}}-G_{{\rm RP}})^2$.
\end{enumerate}

After applying all cuts above, $3{\rm ,}401{\rm ,}809$ stars remain as an initial sample set with precise Gaia astrometric/photometric measurements. Those stars are spread over the sky with a uniform distribution (see Figure~\ref{fig:fig1}), and the absence of any significant over-density at low Galactic latitude confirms that objects in that subset are very local disk/halo stars. 

\subsection{Identification of Halo Candidates from the Reduced Proper Motion Diagram}\label{sec:rpmdiagram}
The kinematics of local halo stars is very distinct and can be used to separate them from the local disk populations. Typical halo stars have high Galactic rotational velocities, $V$, relative to the local standard of rest, $V \ge 200\,{\rm km\,s}^{-1}$ \citep{bensby:14}. Radial velocities for stars are essential to precisely calculate the rotational velocities. However, current Gaia EDR3 provides radial velocities mostly for bright stars. As full radial velocity measurements for all Gaia stars will not be published before 2022, we approach the selection process for halo stars from a different angle by examining the distribution of stars in a reduced proper motion (RPM) diagram. The RPM diagram \citep{luyten:22, jones:72, lepine:05a, kim:20, koppelman:21} is an efficient tool for separating stellar populations. The RPM diagram is distance independent, and efficiently separates giants from main-sequence stars and white dwarfs, but also the RPM diagram segregates stars by transverse motion, which is useful for selecting high-velocity stars from the local halo. 

An RPM diagram plots a reduced proper motion as a function of a star's colour. For stars in the Gaia catalogue, we define a reduced proper motion, $H_G = G + 5 \log\mu + 5$, and from the relationship between the proper motion, parallax, and transverse motion, we find that: $H_G = M_G + 5 \log{\vt} + 1.621$, where $M_G$ is the absolute magnitude of the star, and $\vt$ is the star's transverse motion  in km s$^{-1}$. Values of $H_G$ only require proper motions to be available and accurate.

One caveat of the RPM diagram is that stars in the Galactic plane, especially near the Galactic centre, can be significantly affected by reddening effects due to the amount of gas and dust in the plane of the Milky Way. Since we do not limit our sample to distance or galactic latitude, the line-of-sight extinction value might play an important role in the selection process, even though \citet{andrae:18} report that reddening of stars in the higher Galactic latitudes ($|b| > 50^{\circ}$) is prominently less significant than that of stars in the Galactic plane. 

We adapt the 3D reddening map from \citet{green:19} to obtain colour excess values $E(B-V)$ for each of our stars. To convert the colour excess to Gaia passbands, we referred to \citet{wang:19}, who provides the colour excess ratio and extinction coefficient for Gaia passbands based on the collected data from Gaia, APSS, SDSS, Pan-STARRS DR1, Two-MASS (2MASS), and WISE surveys. From Table 3 in \citet{wang:19}, we obtain the relative extinction values for Gaia passbands, which are:
\begin{equation}
\begin{split}
    A_{V} = 3.16\,E(B - V) =&\,2.394\,E(G_{\mathrm{BP}} - G_{\mathrm{RP}}) \\
    \therefore E(G_{\mathrm{BP}} - G_{\mathrm{RP}}) =&\,1.320\,E(B - V) \\
\end{split}
\end{equation}

\begin{equation}
\begin{split}
    A_{G} = 1.890\,E(G_{\mathrm{BP}} - G_{\mathrm{RP}}),&\,\, A_{G_{\mathrm{RP}}} = 1.429\,E(G_{\mathrm{BP}} - G_{\mathrm{RP}}) \\
    \therefore E(G - G_{\mathrm{RP}}) = A_{G} - A_{G_{\mathrm{RP}}}& = 0.461\,E(G_{\mathrm{BP}} - G_{\mathrm{RP}}) \\
\end{split}
\end{equation}

With those extinction corrections, we redefine the reduced proper motion to be: $H_G = G + 5 \log\mu + 5 - A_G$, which provides an extinction corrected RPM diagram. Since \citet{green:19} constructed their 3D extinction map based on Pan-STARRS photometry, the map does not provide extinction estimates for stars in the Southern hemisphere. Therefore, we assign `$0$' to the colour excess and extinction value for stars in $b \le -30^{\circ}$. For stars in $b > -30^{\circ}$, on the other hand, we obtain the $\grp$ colour excess and $G$ band extinction values based on the equations above. 

\begin{figure*}
\centering
\includegraphics[scale=0.98]{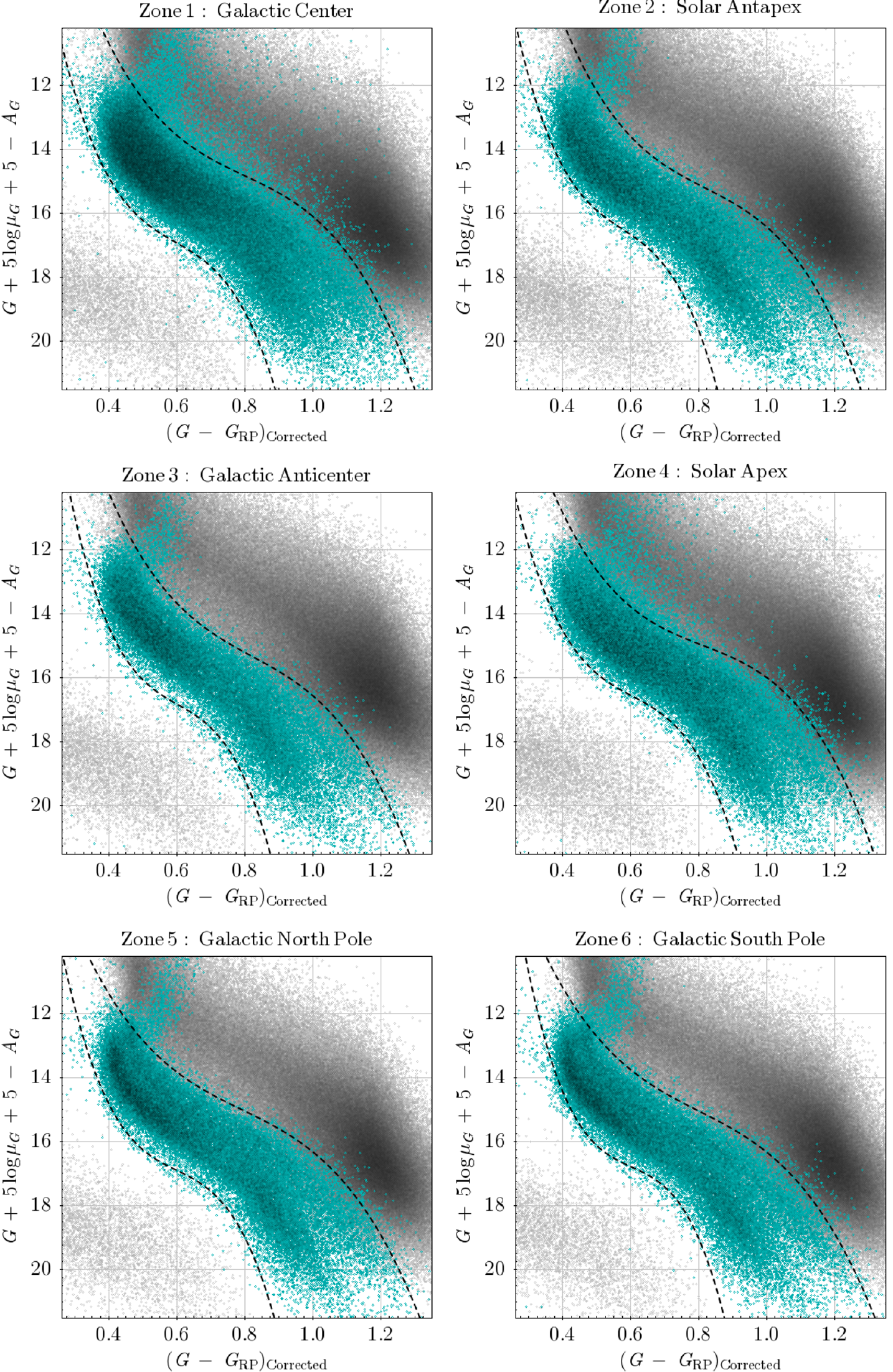}
\caption{Reduced proper motion (RPM) diagrams for high-proper-motion ($\mu > 40$ mas yr$^{-1}$) stars in the Gaia EDR3 in six different regions in the sky (see Figure~\ref{fig:fig1}). High velocity stars ($\vt > 200$ km s$^{-1}$) are plotted as blue dots and identify the general locus of halo stars in each diagram. We define empirical colour-RPM relationships (upper black-dashed lines) to identify all likely halo candidates. We identify $551,214$ stars in total that are selected to be the halo stars. \label{fig:fig2}}
\end{figure*}

As $\sim 30$\% of the high-proper-motion stars are found in the Southern hemisphere, reddening effects remain an issue for our RPM selection. We check how reddening effects influence the selection process of halo main-sequence stars by comparing RPM diagrams of high-proper-motion stars in six different zones of the sky (see Figure~\ref{fig:fig2}). We use $\grp$ colour, instead of using $\bpmrp$ colour, because a) the recent Gaia EDR3 paper \citep{fabricius:20} recommends to use $\grp$ colour if the sample contains faint red sources due to a strong bias for those sources in $G_{{\rm BP}}$ magnitudes, and b) $\grp$ colour is inherently less sensitive to reddening compared to the $\bpmrp$ colour. 

The stellar distribution in each panel in Figure~\ref{fig:fig2} show the typical ``split loci" of the disk and halo main-sequence stars, along with the sparser white dwarf locus on the lower-left corner of the diagram. The halo locus, in the centre of the diagram, can be recognised by its rounded blue end, which consists of old main-sequence turnoff stars. The disk locus is shifted up due to disk stars having lower transverse motions. Subgiants from the halo overlap with solar-type stars in the disk around $H_G \sim 11.0$ and $G - G_{\mathrm{RP}} \sim 0.5$. These diagram do not show the red giants and blue/young main-sequence stars of the disk populations, which have $H_G < 9$.

Comparing the RPM diagrams from the six different directions on the sky shows subtle differences. Some of these are due to colour shift (reddening effect). While stars near both Galactic poles have less scatter than those in the Galactic plane, stars near the Galactic centre show the most dispersed features, which is mostly caused by severe line-of-sight interstellar reddening toward the Galactic centre. If reddening effects are supposed to be small the other reason can be explained by the slightly different kinematics of the stars in different parts of the sky. In the direction parallel to the Sun's motion, the transverse motions of the halo stars are larger or smaller because we are more sensitive to the ``$V$" values, which are larger for the halo stars. This can slightly ``shift" the mean location of the halo sequence depending on the area of interest where we observe.

To select halo stars in the RPM diagram we define arbitrary cuts in the ($\grp, H_G$) plane that are meant to select the vast majority of stars with high-tangential-velocities ($\vt > 200 \,\,\kms$; blue points in Figure~\ref{fig:fig2}), which is an adopted criterion in \citet{gaia:18a} to make a star a likely member of the halo population. Reduced proper motion cuts, however, are more inclusive than pure tangential velocity cuts because (1) they do not rely on the Gaia parallax values, and (2) reduced proper motion cuts separate out halo and disk stars both by velocity and by metallicity as demonstrated in \S~\ref{sec:data} and ~\ref{sec:calibrator} below. Black dashed lines in each RPM diagram are empirical colour-RPM boundaries, which separate the halo population from the disk population and the white dwarfs, based on the general shape of their loci in the RPM diagram. We firstly remove most white dwarfs that have bluer colours and larger RPM values by using an empirical cut which separates white dwarfs and main-sequence stars \citep{kim:20}: $H_{G} \le 4.94(\bpmrp)+12.91$. To acquire a better fit, we limit Gaia colour in the range of $0.2 < \grp$ and RPM value in the range of $11.5 < H_G$. Finally, the lines are defined from a third-degree polynomial fit of the bluest and reddest median $\grp$ colours where the number of high-tangential-velocity stars in $\grp$ colour histograms in each RPM bins is larger than $20$. We use polynomials of the form: $H_{G} = c_0 + c_1(\grp) + c_2(\grp)^2 + c_3(\grp)^3$. The coefficients of the ``blue" and ``red" limits are slightly different based on the zones and are listed in Table~\ref{tab:tab1}.

\begin{table}
\centering
\caption{Coefficients of the polynomials of the form: $H_{G} = c_0 + c_1(\grp) + c_2(\grp)^2 + c_3(\grp)^3$ used as boundaries for the selection of halo main-sequence stars in the RPM diagram.}
\label{tab:tab1}
\resizebox{\columnwidth}{!}{%
\begin{tabular}{lccccc}
\hline
Zone & colour & $c_0$ & $c_1$ & $c_2$ & $c_3$ \\
\hline
\hline
1: Galactic centre & Blue & $- 9.994$ & $123.5$ & $-197.3$ & $110.5$ \\
& Red & $-6.169$ & $68.77$ & $-79.74$ & $33.23$ \\
2: Solar Antapex & Blue & $-13.43$ & $142.5$ & $-236.1$ & $137.2$ \\
& Red & $-11.06$ & $85.34$ & $-97.50$ & $39.64$ \\
3: Galactic Anti-centre& Blue & $-18.67$ & $165.3$ & $-265.2$ & $147.2$ \\
& Red & $-9.149$ & $77.89$ & $-87.77$ & $35.58$ \\
4: Solar Apex & Blue & $-10.11$ & $121.4$ & $-191.9$ & $106.0$ \\
& Red & $-6.988$ & $72.94$ & $-84.69$ & $34.73$ \\
5: Galactic North Pole & Blue & $-14.46$ & $146.8$ & $-236.5$ & $131.6$ \\
& Red & $-2.199$ & $54.86$ & $-62.77$ & $26.42$ \\
6: Galactic South Pole & Blue & $-22.42$ & $185.8$ & $-300.6$ & $166.3$ \\
& Red & $-2.789$ & $56.07$ & $-63.30$ & $26.34$ \\
\hline
\end{tabular}%
}
\end{table}

\citet{gaia:18a} reported bimodal main sequences for stars with high-tangential velocities in the CMD, and our sample also shows this bimodality in the RPM diagram. The double sequences are noticeable in all diagrams, regardless of any line-of-sight direction of the sky. We believe the upper sequence shows the high-velocity tail of the thick disk population; \citet{amarante:20} reported that kinematics of $\sim13$\%~of the high transverse velocity ($v_{T} > 200$ km s$^{-1}$) stars in their model are consistent to that of stars in the thick disk.

\subsection{Collective Calibration Dataset}
To derive photometric metallicities for all the halo stars selected above, we first assemble a list of ``metallicity calibrators" which are stars in our catalogue for which spectroscopic metallicity measurements exist in the literature. A large majority of stars in the halo population is expected to consist of the oldest objects in the Galaxy. Stars in the halo are also expected to be significantly more metal-poor than stars in the disk, which has been confirmed in various spectroscopic studies \citep[and see additional references therein]{carollo:07, carollo:10, an:13, liu:18, naidu:20}. The primary targets in those studies are mostly red giants which allow us to study distant stellar structures extending in the far outer halo. Halo main-sequence stars, which tend to be more local, are thus often overlooked. However, there still exists spectroscopic metallicity measurements ([Fe/H])\footnote{Technically, [Fe/H] should be referred as iron abundance. In this study, however, we define it as stellar metallicity and treat it the same as the overall metal abundance, [M/H].} for small but representative subsets of our halo candidates from various surveys, which are sufficient to define a photometric metallicity grid for low-mass stars (K and M dwarfs/subdwarfs) in the CMD. The datasets we collected are from:
\begin{itemize}
    \item SDSS SEGUE I/II \citep{yanny:09}
    \item SDSS APOGEE DR16 \citep{majewski:17,ahumada:20}
    \item LAMOST DR6 \citep{cui:12}
    \begin{itemize}
        \item Low-resolution A, F, G, and K catalogue
        \item Low-resolution M dwarf catalogue
        \item Mid-resolution catalogue
    \end{itemize}
    \item GALAH DR3 \citep{desilva:15, buder:20}
    \item A low-and-mid resolution catalogue of nearby M dwarfs \citep{hejazi:20}
    \item SDSS M dwarf catalogue\footnote{A full catalogue is provided by S. L{\'e}pine (private communication).}
\end{itemize}

The total number of calibrators from these sources which are found in our catalogue is $20,047$. Each of the subsets is described in more detail below.

\begin{figure*}
\includegraphics[width=\textwidth]{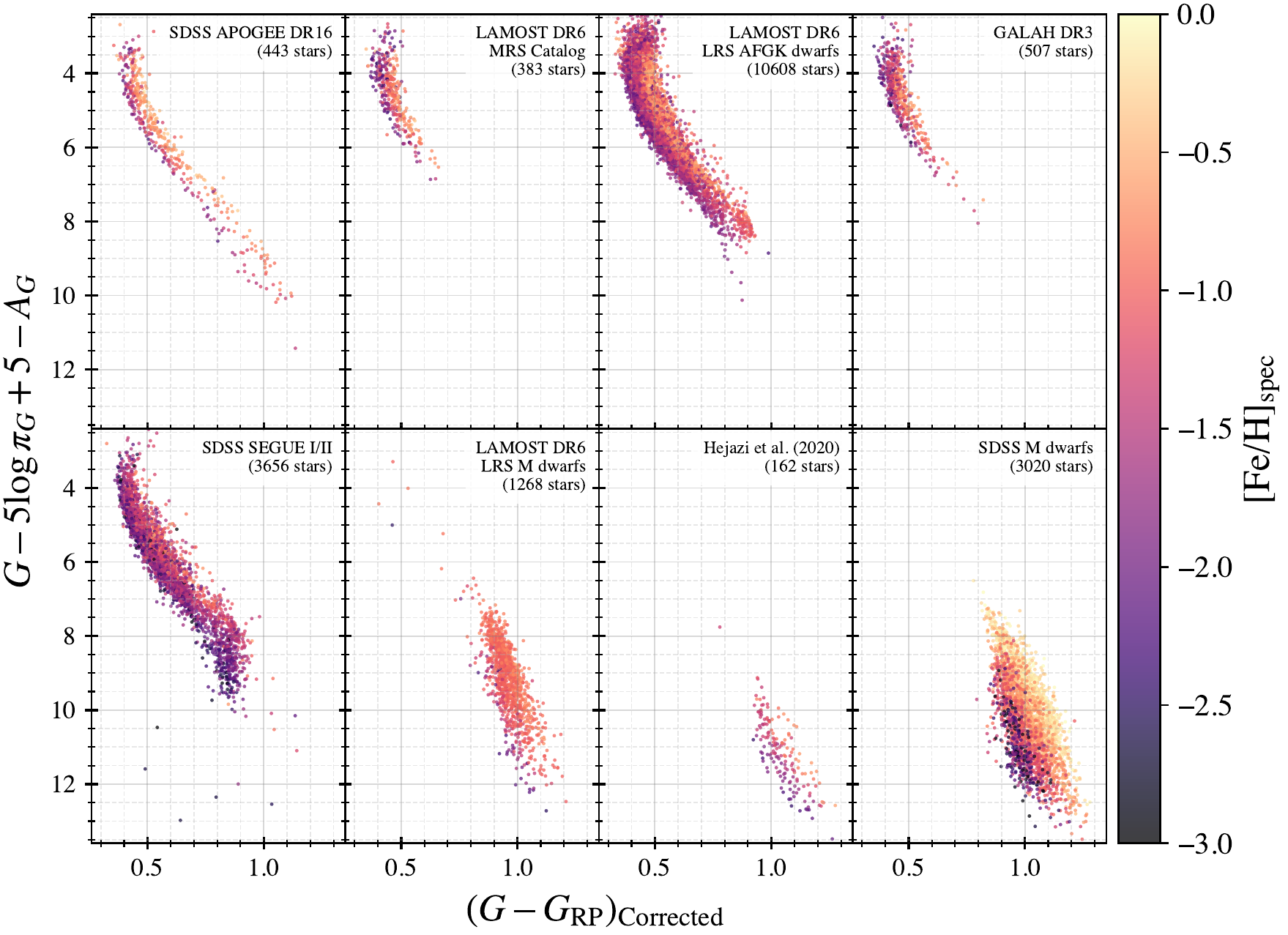}
\caption{CMDs of stars with precise spectroscopic metallicities ($\sigma({\rm [Fe/H]})/{\rm [Fe/H]} \le 0.15$) from SDSS APOGEE DR16 (443 stars), LAMOST DR6 MRS (383 stars) and LRS catalogues (10,608 stars from the AFGK dwarfs catalogue and 1,268 stars from the M dwarf catalogue), GALAH DR3 (507 stars), SDSS SEGUE-I/II (3,656 stars), \citet{hejazi:20} (162 stars), and SDSS M dwarf catalogue (3,020 stars). For convenience, we use [Fe/H], instead of [M/H], for describing chemical abundances of M dwarfs. For stars in the SDSS M dwarfs catalogue, we uniformly assigned random [Fe/H] values depending on their subtypes (also see Table~\ref{tab:tab2}).  \label{fig:fig3}}
\end{figure*}

\subsubsection{SEGUE-I/II and APOGEE DR16}
One of the best-known, all-sky spectroscopic survey for stars in the Galaxy is the Sloan Extension of Galactic Understanding and Exploration survey \citep[SEGUE; ][]{yanny:09} and the Apache Point Observatory Galactic Evolution Experiment \citep[APOGEE; ][]{majewski:17,ahumada:20} survey. While the SEGUE survey is a moderate-resolution ($R \sim 1,800$) spectroscopic survey in optical wavelength for faint stars ($14.0 < g < 20.3$), the APOGEE survey is a high-resolution ($R \sim 22,500$), high signal-to-noise survey in infrared wavelength with a focus on red giants but including many main-sequence stars as well. We cross-matched our catalogue with SDSS SEGUE-I/II and APOGEE DR16, then removed stars with large uncertainties in stellar abundances ($\sigma({\rm [Fe/H]})/{\rm [Fe/H]} \ge 0.15$) and a SEGUE star with solar metallicity (${\rm [Fe/H]} > 0.0$). There remained 443 stars from APOGEE and 3656 stars from SEGUE. Top-left and bottom-left panels in Figure~\ref{fig:fig3} show their distribution in the CMD, with their metallicity values encoded on a colour scale. 

\subsubsection{LAMOST DR6}
The Large Sky Area Multi-Object Fiber Spectroscopic Telescope \citep[LAMOST;][]{cui:12} has been conducting low-resolution ($R \sim 1,800$) and moderate-resolution ($R \sim 7,500$) spectroscopic surveys in optical wavelength. We cross-matched our catalogue with the low-resolution A, F, G, and K star (LRS-AFGK) catalogue, the low-resolution M star (LRS-M) catalogue, and the mid-resolution parameter (MRS) catalogue in LAMOST DR6\footnote{\url{http://dr6.lamost.org/v2/}}. After removing stars with large uncertainties ($\sigma({\rm [M/H]})/{\rm [M/H]} \ge 0.15$) in their metallicities\footnote{M dwarf catalogue provides overall metal abundance, [M/H], which is technically different with Fe abundance, [Fe/H]. In this study, however, we treat both terms the same.} and five LRS-AFGK stars with apparent solar metallicities, we retained $10,608$ stars from the LRS-AFGK, $1,268$ stars from the LRS-M, and $383$~stars from the MRS. The second and third panels in the top row and the second panel in the bottom row of Figure~\ref{fig:fig3} show the distribution of stars in the CMD from MRS catalogue, LRS-AFGK, and LRS-M, respectively, with their metallicity encoded on the colour scale.

\subsubsection{GALAH DR3}
The recent GALactic Archaeology with HERMES (GALAH) DR3 provides stellar parameters and elemental abundances for $588,571$ stars that have been observed with the HERMES ($R \sim 28,000$) spectrograph \citep{desilva:15, buder:20}. The database contains $507$ matches with our catalogue. All matches appear to be intermediate mass stars (A, F, G type) near the main-sequence turnoff, and their distribution in the CMD is shown in the top-right panel in Figure~\ref{fig:fig3}, again with metallicity values encoded by colour.

\subsubsection{\citet{hejazi:20}}
All the catalogues above include very few late-type M stars, which are critical to our calibration of the colour-magnitude-metallicity relationship at the low-mass end. For that purpose, we cross-match with a recent spectroscopic catalogue of nearby, high-proper-motion M dwarfs and subdwarfs \citep{hejazi:20}. The catalogue contains metallicities ([M/H]) obtained from low-and-mid resolution ($2,000 \le R \le 4,000$) optical spectroscopic observation. After cross-matching the positions on the sky, we exclude the stars with large uncertainties in [M/H], which leaves $162$ stars. The third panel in the bottom row of Figure~\ref{fig:fig3} shows the distribution of stars in the CMD with their metallicity colour-coded.

\subsubsection{SDSS M dwarf catalogue}

\begin{table}
\caption{Metallicity boundaries for stellar subtypes.}
\label{tab:tab2}
\resizebox{\columnwidth}{!}{%
\begin{tabular*}{\columnwidth}{l@{\extracolsep{\fill}}c}
\hline
Subtype & [Fe/H] \\
\hline
\hline
M  & $-0.34$ to $0.00^*$ \\
sdM & $-0.87$ to $-0.34$ \\
esdM & $-1.36$ to $-0.87$ \\
usdM & $-3.00^*$ to $-1.36$ \\
\hline
\end{tabular*}}
\footnotetext{*}{$^*$ We limited the maximum and minimum values of SDSS M dwarf candidates, considering metallicity ranges for other dataset.}\\ 
\end{table}

Due to the relative paucity of low-mass stars in our calibration subset, we add information for an additional $3,020$ M dwarfs/subdwarfs identified in the SDSS spectroscopic archive. While these stars do not have formal metallicity measurements, we can extract crude metallicity estimates from their spectral subtype. We obtain spectral types by classifying the stars against classification templates using the method described in \citet{zhong:15}. We use the classification of the stars as dwarf (dM), subdwarf (sdM), extreme-subdwarf(esdM), and ultra-subdwarf(usdM), and assigned uniformly distributed random values in between the metallicity ranges for each stellar subtypes, defined from the calibrated relationship between molecular indices, $\zeta_{\mathrm{TiO/CaH}}$, and metallicities \citep{woolf:09}. The metallicity ranges for each subtype are provided in Table~\ref{tab:tab2}. We show the distribution of these stars in the CMD in the bottom-right panel in Figure~\ref{fig:fig3}.

\section{Photometric Metallicity Estimates for the Halo Candidates}\label{sec:phot_feh}
The left panel in Figure~\ref{fig:fig4} shows the full set of $20,047$~stars in the collective dataset with stellar metallicities in the range of $-3.0 \le {\rm [Fe/H]} < 0.0$. Although only $3.6$\% of stars from our full catalogue of halo candidates have spectroscopic metallicities, their distribution spans a wide range of metallicity values along the entire main-sequence, and notably shows a clear metallicity trend which suggests a linear relationship between metallicity and $\grp$ colour at a given absolute magnitude. Using this distribution we create a calibration grid from which to estimate metallicities for all the stars in our catalogue of halo candidates. We describe the multi-step procedure in detail below.

\subsection{Duplicates and Removal of Unresolved Binaries}\label{sec:calibrator}
\begin{figure*}
\centering
\includegraphics[width=\textwidth]{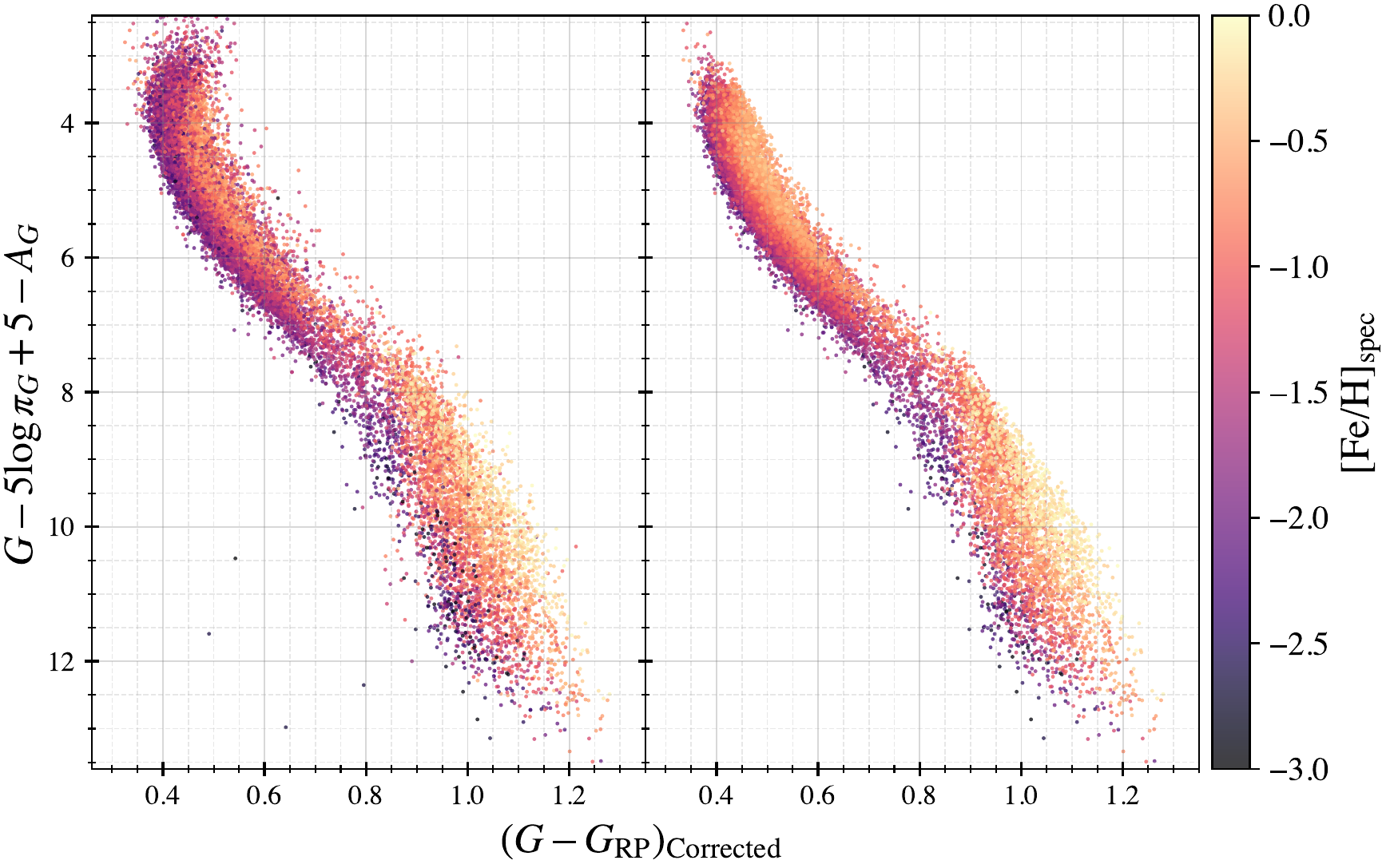}
\caption{CMDs for all stars ($20,047$ stars; left) in the collective calibration dataset and for a ``cleaned" subset ($17,170$ stars; right) obtained after removing duplicates and contaminants. We assign weighted averages to spectroscopic metallicities of $856$ stars that have been observed in multiple surveys (duplicates), then remove stars most likely to be unresolved pairs and WD+M binaries, using the procedure described in \S~\ref{sec:calibrator}. Most turn-off stars end up being eliminated as well from the binary removal procedure. \label{fig:fig4}}
\end{figure*}

Stars shown in the left panel in Figure~\ref{fig:fig4} include $856$ stars that have been observed multiple times by two or three different spectroscopic surveys. For those stars, we calculate weighted means and assign them as their spectroscopic metallicities. While calculating the weighted means, we exclude values from the SDSS M dwarf catalogue as those values are not actual spectroscopic metallicity measurement.

We also remove stars that appear to be unresolved binary systems; these stars are bad calibrators for the colour-luminosity-metallicity grid since their composite colours and luminosities are not a reliable indicator of metal content. To remove binaries, we firstly group the calibrators in 13 different metallicity bins; the first and last groups include all stars in the metallicity range of ${\rm [Fe/H]} < -2.4$ and $-0.2 \le {\rm [Fe/H]}$, respectively. Other groups are composed of stars in the range of $-2.4 \le {\rm [Fe/H]} < -0.2$ in $0.2$ dex increments. For each group, we iteratively fit a fourth-degree polynomial to the stellar distribution in the CMD, and for each star we calculate the absolute $G$ magnitude difference ($\Delta M_{G}$) from the fit. For each of the first three iterations, an outlier is removed if its $\Delta M_{G}$ is more than $5\sigma$ away from the mean of the overall $\Delta M_{G}$ distribution. For the last two iterations, we add an additional condition that removes any overluminous star with $\Delta M_{G} \ge 0.7$. 

The right panel in Figure~\ref{fig:fig4} shows the final distribution of the cleaned collective dataset ($17,170$ stars) in the CMD. The few WD+M pairs below the main sequence and most of unresolved binaries are expected to have been eliminated from the subset. The cleaning process yields a much smoother distribution of metallicity values, and also eliminates most of the evolved stars with $M_G < 3.0$ (which are also technically over-luminous).

\subsection{Metallicity Estimates Using K-Nearest neighbour Regressor}

\begin{figure}
\centering
\includegraphics[width=\columnwidth]{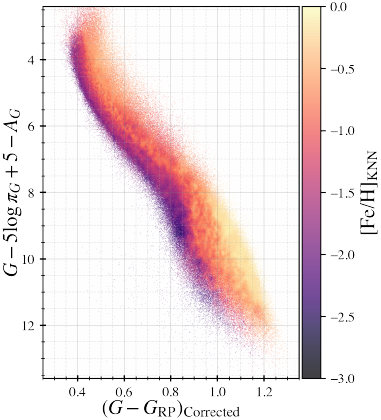}
\caption{A CMD of all halo candidates with their photometric metallicity estimated from the KNN regressor. This provides a crude representation of the colour-magnitude-metallicity relationship for these low mass stars, but irregularities in the estimated metallicity distribution suggests random and systematic errors in the KNN calibrators, and possible effects due to unresolved binaries. \label{fig:fig5}}
\end{figure}

To verify that our calibration set spans the full range of the colour-magnitude distribution of our full catalogue of halo candidates, and to identify other possible shortcomings, we obtain crude metallicity values from the Python package, \texttt{Scikit-learn} \citep{pedregosa:11}, which performs a {\it K-Nearest neighbour} (KNN) Regressor fit (\texttt{sklearn.neighbors.KNeighbors\\Regressor}) with a neighbour distance of $k = 20$ and `distance' weights. Figure~\ref{fig:fig5} shows a CMD for all halo candidates with their metallicity estimates from the KNN regressor in the metallicity range of $-3.0 \le {\rm [Fe/H]_{KNN}} \le 0.0$. 

The overall metallicity distribution appears to follow the trends of the clean collective calibration dataset, however, there are few artificial features that come in sight. One would normally expect that the mean metallicity of the stars should be similar up and down the main-sequence, however stars with $0.75 \le \grp \le 0.85$ and $8.5 \le M_{G} \le 10.0$ appear to have their metallicity underestimated by the model, probably because the sources from the collective dataset in that area are mostly stars flagged to be very metal-poor by the SDSS SEGUE. In addition, there is a metallicity disconnect in the lower main sequence where early M subdwarfs are located ($0.85 \le \grp \le 0.95$, $9.5 \le M_{G} \le 11.0$) where most stars are flagged by the KNN regressor as being more metal-rich than stars found elsewhere on the main-sequence. We believe that this model overestimates metallicities for stars in that area in part due to the lack of metal-poor stars in the area of the collective dataset (see Figure~\ref{fig:fig5}). However, it also appears likely, based on the shift in the mean metallicity along the main-sequence, that the spectroscopic metallicity estimates are themselves affected by significant systematic errors, which tend to register early M dwarfs as more metal-poor than what they really are, and late-M dwarf as more metal-rich. There are also many small patches, mainly found in the A, F, G, and K dwarfs ranges, which shows smoothing issues even though we selected $20$ neighbours to estimate the metallicity values. One possible reason for this is the incomplete elimination of unresolved binaries, which might still contaminate the calibration sample.

\begin{figure}
\centering
\includegraphics[width=\columnwidth]{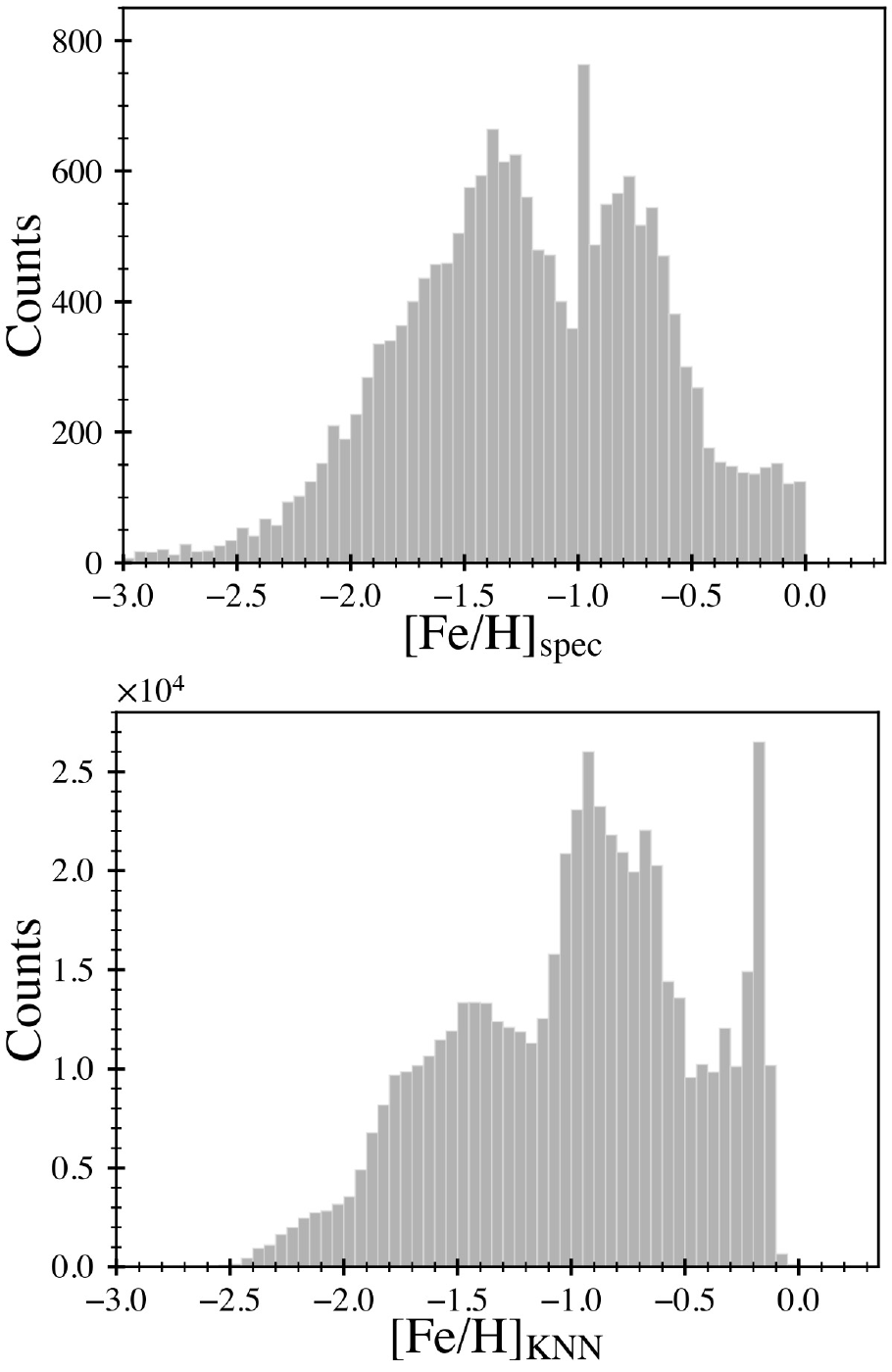}
\caption{Top: histogram of spectroscopic metallicities for the $17,170$ stars in the collective dataset. Bottom: histogram of metallicity estimates from the KNN regressor fit for halo candidates. The overall shape of both histograms appears similar. A bimodal peak in both distributions reflects the signature of the thick disk and halo populations. \label{fig:fig6}}
\end{figure}

A recent study on the metallicity distribution of high-tangential-velocity stars \citep{gaia:18a} reported that their metallicity distribution shows double peaks, with peak metallicities of $-1.3$ and $-0.5$ dex, and suggest that each peak corresponds to the mean metallicity of the halo and thick disk population, respectively. \citet{gaia:21} supports this idea by the fact that a \texttt{PARSEC} isochrone with metallicity of $-0.5$ dex and age of 11 Gyr aligns with the gap between the double sequences, even though the isochrone has to be shifted by 0.04 in $\bpmrp$ colour and 0.2 in $G$ magnitude. 

Histograms of spectroscopic metallicity estimates and the metallicity estimates from the KNN regressor also support the dual peaks of the high-tangential-velocity stars, but with slightly different peak metallicity values. A top panel in Figure~\ref{fig:fig6} shows the histogram of spectroscopic metallicities for the $17,170$ stars in the collective dataset. It shows two major peaks at $-0.75$, and $-1.40$ dex. The narrow peak at $-1.0$ dex is an artifact from the LAMOST DR6 LRS M dwarf catalogue, which appears to have many stars with metallicity values arbitrarily set to $-1.0$. The other two peaks are, however, more likely the true signature of the assumed dual populations of the halo. The bottom panel in Figure~\ref{fig:fig6} shows the histogram of metallicity estimates from the KNN regressor, whose overall shape appears to be similar to that in the top panel. There is a narrow peak at $-0.2$ dex which is likely an artifact. It appears that the two metal-poor peaks in both panels have similar values, both consistent with the hypothesised dual populations, however, the value of the more metal-rich peak in both panels ($\mathrm{[Fe/H]} \simeq - 0.7$), which is supposed to represent the thick-disk population, seem to be more metal-poor than the reported value from other studies. This may suggest that our reduced proper motion selection is more efficient in eliminating the more metal-rich stars of the thick disk population. The many artifacts found in Figure~\ref{fig:fig5} however suggest that one should be cautious, and a more careful calibration method should be used.

\subsection{Metallicity Estimates from a Calibrated Photometric Metallicity Grid}
Although the overall shape of the distribution of metallicity estimates from the KNN regressor generally follows the results from the large spectroscopic surveys and could provide a crude estimate of a star's metallicity, estimates of a number of stars (especially M dwarfs) appear to be under-/over-estimated by the regressor. To improve our metallicity estimates, henceforth, we define a photometric metallicity grid in the space of $\grp$ colour and $M_{G}$, whose shape we define using stars in the collective dataset.

\subsubsection{Building the Photometric Metallicity grid}\label{sec:phot_grid}
\begin{figure*}
\centering
\includegraphics[width=\textwidth]{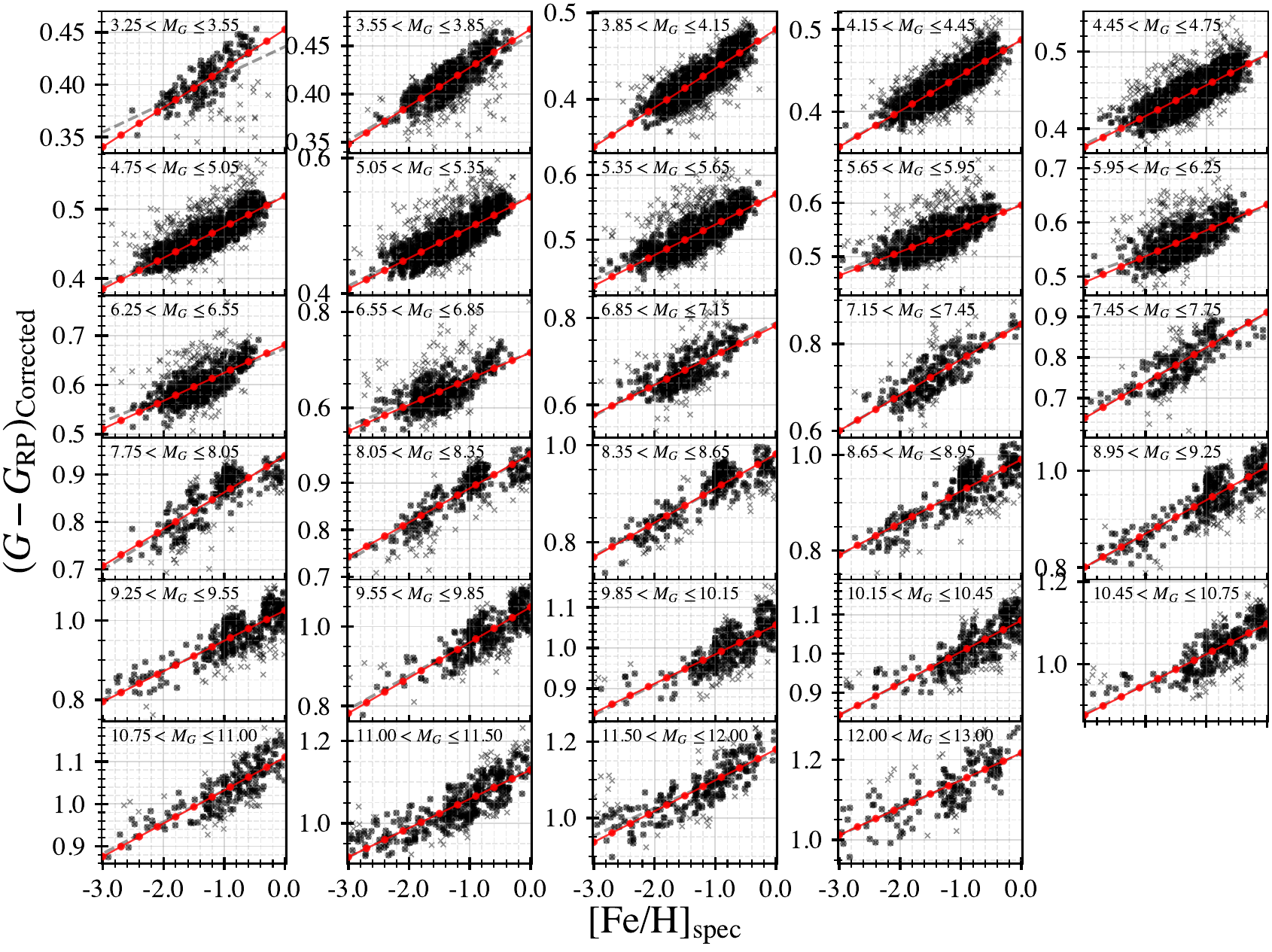}
\caption{Linear fits to stars in each bin of absolute magnitude $M_{G}$ in the parameter space of spectroscopic metallicity, [Fe/H]$_{\mathrm{spec}}$, and $\grp$ colour. Grey dashed lines are the initial linear fits to data and red solid lines are the final fits to black points that have selected after cleaning the outliers (grey crosses). Red points are the initial grid points of $\grp$ colours as given [Fe/H] values in each $M_{G}$ bin. \label{fig:fig7}}
\end{figure*}

We grouped $17,170$ stars in the collective dataset in bins of absolute magnitude in the range of $3.25 \le M_{G} < 11.0$ in $0.3$ mag increments, in the range of $11.0 \le M_{G} < 12.0$ in 0.5 mag increments, and in the range of $12.0 \le M_{G} < 14.0$ in $1.0$ mag increments. The increasing size of the magnitude increments at fainter absolute magnitude ensures that significant number of stars are found in every bin. As shown in Figure~\ref{fig:fig7}, we fit linear relationships between $\grp$ colours and metallicities for stars in each magnitude bin, and then apply the fitted relationships to evaluate the $\grp$ colours corresponding to $12$ fixed metallicity values, ranging from $-3.0$ to $0.3$ dex, in $0.3$ dex increments. Considering that the spectroscopic metallicity of stars in the collective datasets range between $-3.0$ and $0.0$ dex, $\grp$ values corresponding to [Fe/H] = $0.3$ technically represents an extrapolation of the fitted linear relationships. The extrapolated grid line of [Fe/H] = $0.3$ allows us to cover the significant number of more metal-rich stars that lie above the solar metallicity grid; we did not see the need to extrapolate a grid line at the low-metallicity end of the distribution given the very small number of sources found below the grid line of [Fe/H] = $-3.0$. Due to significant scatter in the distribution (likely due in part to measurement inaccuracies but also systematic errors from, e.g. unresolved binaries) we obtain a stronger fit by iteratively excluding outliers. We go through six iterations of the fit, excluding outliers with their differences from the fit higher than $5\sigma$ (for first three runs), $4\sigma$ (for next two runs), and $3\sigma$ (for the last run). The dashed lines in Figure~\ref{fig:fig7} show the initial fits to the data, rejected outliers are represented as grey dots, and the red linear lines are the finalised fits to the retained data points that are shown as dark symbols.

\begin{figure}
\centering
\includegraphics[width=\columnwidth]{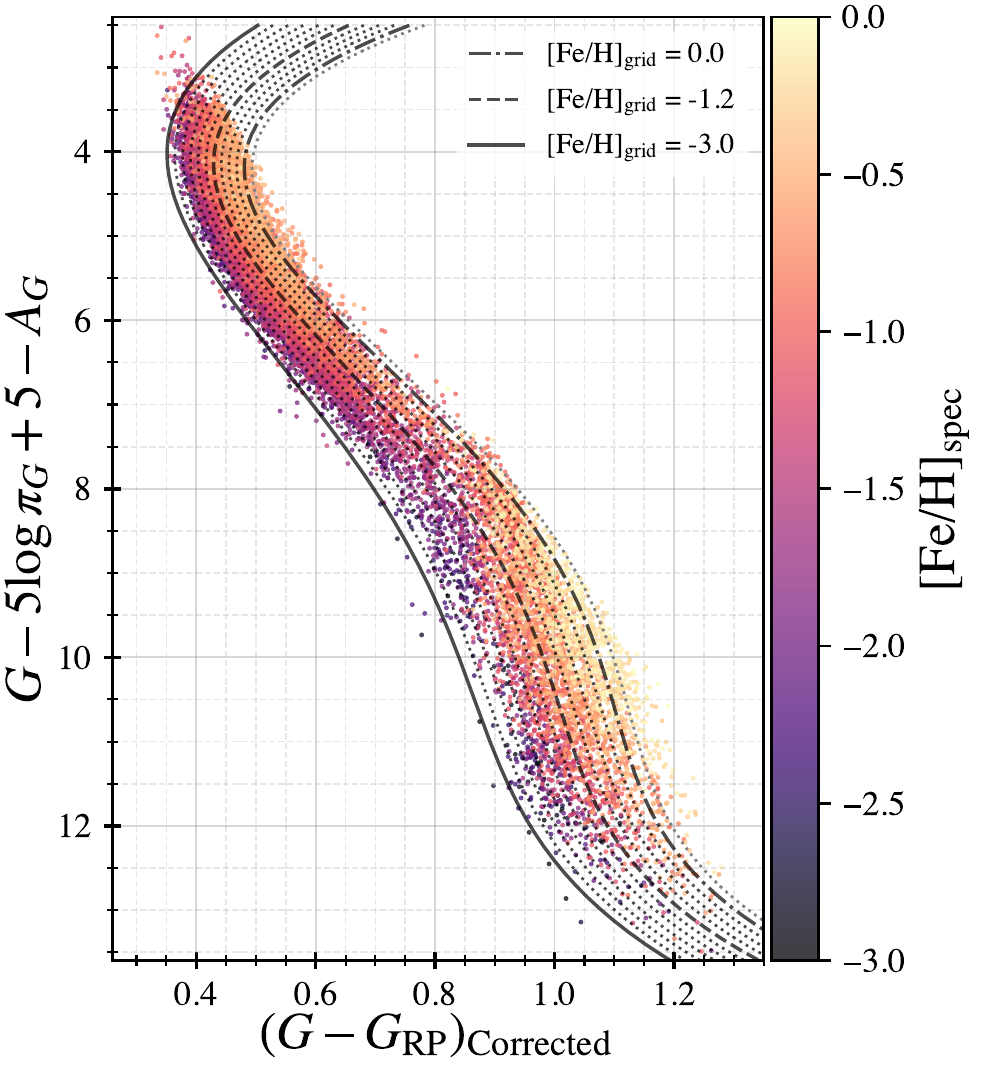}
\caption{CMD of stars in the collective calibration dataset with spectroscopic metallicities represented on the colour scale, and with the resulting photometric metallicity grid overlaid. The grid lines have assigned [Fe/H] values that vary from −3.0 (left) to +0.3 (right), in 0.3 dex increments. Dash-dotted, dashed, and solid lines represent the grid lines with metallicity values, $\mathrm{[Fe/H]} = 0.0, -1.2$, and $-3.0$, respectively. A grey dotted line shows the extrapolated grid with the metallicity of $\mathrm{[Fe/H]} = 0.3$. \label{fig:fig8}}
\end{figure}

We then generate the set of ($\grp$) colours estimated from the linear fits for each pre-defined metallicity value, along with the median $M_G$ values for all the stars used in the fit for each bin. For each set of ($\grp$, $M_G$)$_{\mathrm{[Fe/H]}}$ values, we fit a fourth-degree polynomial. Each one of these polynomials defines one grid line, the combination of which defines a grid with metallicities ranging from $-3.0$ to $0.3$ dex, which is the photometric metallicity grid shown in Figure~\ref{fig:fig8} (black lines). To adjust the shape of the grid at the high-mass end, we assign arbitrary weights to absolute magnitudes of certain data points to improve the regularity of the grid. Notably we set a lower weight ($w = 0.01$) for data points in the absolute magnitude range $M_G \le 3.5$, and a higher weight ($w = 2.5$) for data points in the absolute magnitude range $3.9 < M_G \le 6.0$, with all other data points assigned a weight $w=1.0$. Figure~\ref{fig:fig8} shows the $12$ resulting calibrated grid lines. The dash-dotted, dashed, and solid lines are grid lines with the most metal-rich ($\mathrm{[Fe/H]} = 0.0$), the median metallicity ($\mathrm{[Fe/H]} = -1.2$), and the most metal-poor ($\mathrm{[Fe/H]} = -3.0$), respectively. A grey dotted line shows the extra grid line extrapolated to the metallicity value, $\mathrm{[Fe/H]} = 0.3$. 

A metallicity estimate from the grid ($\mathrm{[Fe/H]_{grid}}$) is obtained by interpolating the grid with a given $\grp$ colour and $M_{G}$. Metallicity estimates for stars outside the grid have values set to $-3.0$ or $0.3$ dex which are the lowest and highest metallicity limits of our grid. Metallicity estimates for stars within the grid limits, on the other hand, are interpolated from the grid, even though the grid line for [Fe/H] = $0.3$ is linearly extrapolated from the $12$ grid lines.

\subsubsection{Vetting and Recalibration using Common-proper-motion Pairs}\label{sec:binaries} 
Our catalogue of halo candidates happens to contain a significant number of common proper motion pairs, most of which are likely  to be wide binary systems. Wide binaries are assumed to have formed in the same star-forming environment, and thus are expected to share the same chemical composition. This allows us to use these pairs to validate the photometric metallicity grid, as a kind of internal consistency check. We crossmatched our halo catalogue to the SUPERWIDE catalogue \citep{hartman:20} and recovered $1\mathrm{,}246$ wide binary candidates with high Bayesian probabilities ($>95\%$) of being physical pairs. If we assume $\mathrm{[Fe/H]_{grid}}$ of pairs are precise, and there is no other unresolved companion in these systems, then the difference in estimated metallicity between the primary and secondary of the common-proper-motion pair should be close to zero. Large metallicity differences between some component stars may indicate the presence of unresolved companions, but systematic offsets, in particular for pairs of different masses, can reveal errors in the calibration of the grid, notably fundamental problems due to systematic errors in the spectroscopic metallicity estimates of the calibration stars. Effects of binarity should be best revealed by examining pairs of stars of similar masses, for example K+K binaries, while more serious systematic errors in the calibration of the grid should be best revealed by comparing pairs of stars of different masses, like K+M binaries.

\begin{figure}
\centering
\includegraphics[width=\columnwidth]{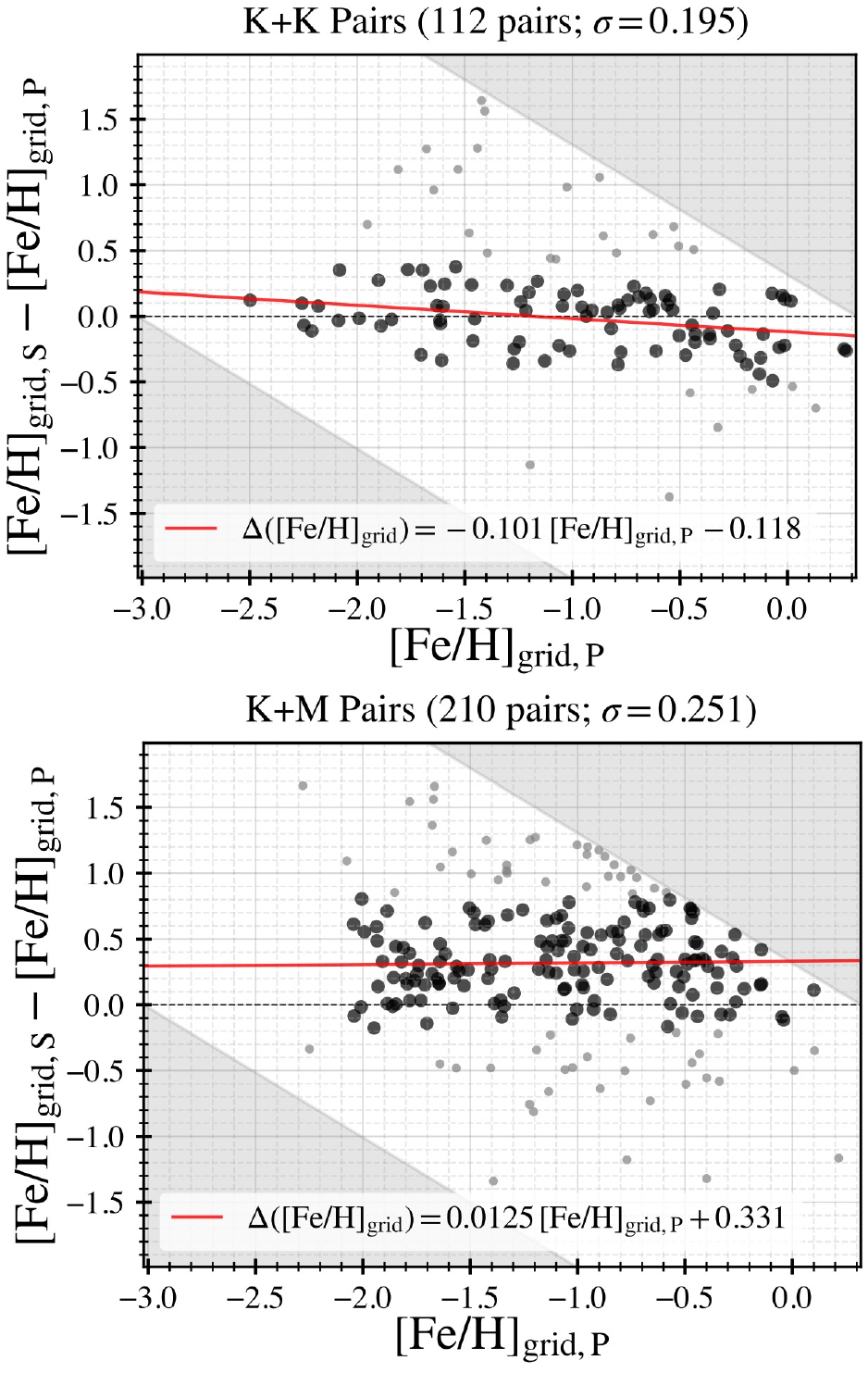}
\caption{Scatter plots displaying the difference in the estimated photometric metallicity values from the grid between the primary and secondary star for 116 K+K (top) and 213 K+M (bottom) common-proper-motion pairs from the SUPERWIDE catalogue, with high Bayesian probability ($\ge 95\%$) of being physical binaries. Red lines are linear fits to stars (black points) whose metallicity deficit is less than $2\sigma$ of the overall distribution after excluding outliers (grey points). Shaded regions at the top-right and bottom-left of the panels represent the metallicity deficit limits set by our grid. \label{fig:fig9}}
\end{figure}

In the top panel of Figure~\ref{fig:fig9}, we plot differences in the photometric metallicity estimates ($\Delta\mathrm{[Fe/H]} = \mathrm{[Fe/H]_{grid, Secondary}} - \mathrm{[Fe/H]_{grid, Primary}}$) between primaries and secondaries for $116$ K+K pairs with metallicity estimates between $-3.0 < \mathrm{[Fe/H]_{grid}} < 0.3$, these are plotted as a function of $\mathrm{[Fe/H]_{grid}}$ of the primaries. Grey shaded regions represent $\Delta\mathrm{[Fe/H]}$ limits set by our metallicity grid. A red solid line shows the fit to $78$ data points (in black), which remain after 10 iterations to exclude $38$ outliers with $\Delta\mathrm{[Fe/H]}$ higher than $2\sigma$ of the distribution; this is done to exclude outliers that may be, e.g., unresolved systems. Due to the $\Delta\mathrm{[Fe/H]}$ limits in both metal-rich and metal-poor ends, we only use data points with $-2.5 < \mathrm{[Fe/H]_{grid}} < 0.0$. The standard deviation, $\sigma = 0.195$, shown in the title is the dispersion of the black points. The overall fit is mostly consistent with zero offset, as we expected, even though the fit shows a weak negative relationship.

The fraction of outliers among the K+K pairs may suggest that $32$\% of halo wide binaries are multiple star systems (i.e., triples or quadruples). This large multiplicity fraction may affect the shape and location of the grid in the CMD. If our calibrators contain many overluminous, unresolved binaries, the grid that is defined by the calibrators could become slightly redder and brighter. The large multiplicity fraction may also explain why there is a large scatter and many outliers in the diagrams shown in figure~\ref{fig:fig7}. Above all else, therefore, a careful selection for K+K binaries by excluding multiple star systems as many as possible is required to create the better calibration fit for the grid.

The bottom panel in Figure~\ref{fig:fig9} shows the same $\Delta\mathrm{[Fe/H]_{grid}}$ distribution, but for K+M pairs. This time we notice not just a larger scatter in the distribution, but also a systematic offset: a linear fit to the distribution of stars with $-2.5 < \mathrm{[Fe/H]_{grid}} < 0.0$ shows an average offset of $\sim 0.33$ dex, which strongly suggests that the metallicities of our M dwarf calibrators are overestimated by about that much. This offset may be due to a different definition of ``metallicity" for K and M stars. Metallicity values for K stars may be more closely tied to Fe abundances. While early-type dwarf spectra have major features that are mostly ionised elemental lines, the major characteristics of M dwarf spectra are the presence of strong absorption bands due to atomic and molecular bands, such as TiO, VO, and CaH, and relatively weak Fe lines. This means that M dwarf metallicities are more closely related to abundances of $\alpha$-elements, and can thus overestimate Fe abundances if halo stars happen to show high [$\alpha$/Fe] ratios. Another possible problem is that many of the ``spectroscopic" metallicity values for our M dwarf calibrators are not derived from detailed spectral analysis but from simple spectral subtypes (i.e., the SDSS M dwarf subset). In any case, what matters is that the observed systematic offset indicates that the colour-magnitude-metallicity grid is not properly calibrated at this step, and some correction may be needed. 

\subsubsection{Grid Recalibration}

\begin{figure*}
\centering
\includegraphics[width=\textwidth]{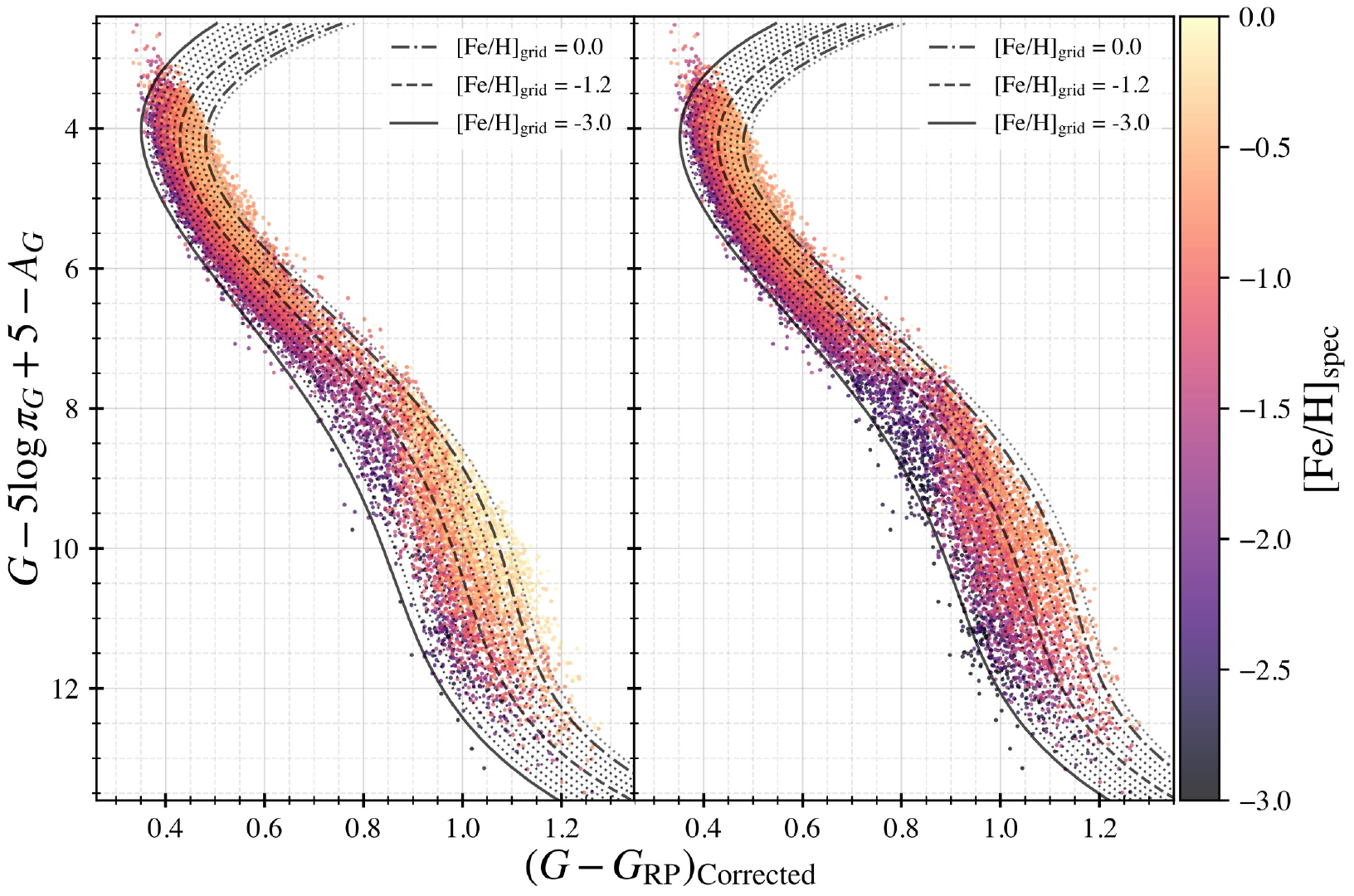}
\caption{Left: Same plot as Figure~\ref{fig:fig8}, showing the initial photometric metallicity grid. Right: metallicity distribution and grid after a corrective offset is applied to the metallicity values of the M dwarfs. Spectroscopic metallicities for stars with $M_{G} > 7.5$ are corrected by subtracting $0.5$ dex to correct for the inconsistency in the metallicity values of K+M pairs, as shown in the bottom panel in Figure~\ref{fig:fig9}. The offset brings the photometric metallicities of K+M pairs in agreement, as shown in the bottom panel of Figure~\ref{fig:fig11}.
\label{fig:fig10}}
\end{figure*}

We will assume that the spectroscopic metallicities of the FGK stars in our calibration sample are accurate, and apply arbitrary corrections to the M dwarf metallicity values to make the grid more self-consistent. After trial and error, we apply a systematic correction to all calibration stars $M_G > 7.5$ by subtracting $0.5$ dex from their initial spectroscopic metallicity values. We then redo all steps we discussed in \S\ref{sec:phot_grid} and derive a new metallicity grid. The offset value of $0.5$ dex is slightly larger than the offset we obtained from the fit, which may appear to be an over-correction, buhe one also needs to consider the fact that an M dwarf bias in the calibration subset could also have dragged off other parts of the grid, especially the nearby K dwarf regime.

Figure~\ref{fig:fig10} shows the previous grid in the left panel and the revised calibrated grid in the right panel. Black lines indicate the most metal-poor (solid line; [Fe/H] = $-3.0$), intermediate (dashed line; [Fe/H] = $-1.2$), and the most metal-rich (dash-dotted line; [Fe/H] = 0.0) metallicities in our grid and dotted lines in between them represent metallicities in range of $-2.7 \le \mathrm{[Fe/H]_{grid}} \le 0.0$. The grey dotted line is the extrapolated metallicity grid for [Fe/H] = $0.3$. The overall shape of the grid appears to be same, however, its coverage of the low-mass ends has been improved, and it no longer looks like the lowest mass stars in our subset are systematically more metal rich than the highest mass ones. Table~\ref{tab:tab3} lists the coefficients of each photometric metallicity grid line as a function of a given $\grp$ colour and $M_{G}$. Each grid line is fit with the polynomial equation:
\begin{equation}
\grp = c_0 + c_1(M_G) + c_2(M_G)^2 + c_3(M_G)^3 +  c_3(M_G)^4
\end{equation}

\begin{table}
\centering
\caption{Coefficients of the polynomials for the finalised photometric metallicity grid.}
\label{tab:tab3}
\resizebox{\columnwidth}{!}{%
\begin{tabular}{cccccc}
\hline
Metallicity & $c_0$ & $c_1$ & $c_2$ & $c_3$ & $c_4$ \\
(dex) & & & & & \\
\hline
\hline
$0.3^{*}$ & $3.218$ & $-1.666$ & $0.3457$ & $-0.02800$ & $0.0008013$ \\
   $0.0$  & $3.123$ & $-1.618$ & $0.3361$ & $-0.02723$ & $0.0007797$ \\
   $-0.3$ & $3.028$ & $-1.570$ & $0.3264$ & $-0.02646$ & $0.0007581$ \\
   $-0.6$ & $2.934$ & $-1.522$ & $0.3168$ & $-0.02569$ & $0.0007365$ \\
   $-0.9$ & $2.839$ & $-1.474$ & $0.3071$ & $-0.02492$ & $0.0007149$ \\
   $-1.2$ & $2.745$ & $-1.426$ & $0.2975$ & $-0.02414$ & $0.0006933$ \\
   $-1.5$ & $2.650$ & $-1.378$ & $0.2878$ & $-0.02337$ & $0.0006717$ \\
   $-1.8$ & $2.556$ & $-1.330$ & $0.2782$ & $-0.02260$ & $0.0006501$ \\
   $-2.1$ & $2.461$ & $-1.282$ & $0.2686$ & $-0.02183$ & $0.0006285$ \\
   $-2.4$ & $2.367$ & $-1.234$ & $0.2589$ & $-0.02105$ & $0.0006069$ \\
   $-2.7$ & $2.272$ & $-1.186$ & $0.2493$ & $-0.02028$ & $0.0005853$ \\
   $-3.0$ & $2.177$ & $-1.139$ & $0.2396$ & $-0.01951$ & $0.0005637$ \\
\hline
\multicolumn{6}{l}{Polynomial equation: $\grp = c_0 + c_1(M_G) + c_2(M_G)^2 + c_3(M_G)^3 +  c_3(M_G)^4$} \\
\multicolumn{6}{l}{$^*$ Extrapolated metallicity grid} \\
\end{tabular}%
}
\end{table}

\begin{figure}
\centering
\includegraphics[width=\columnwidth]{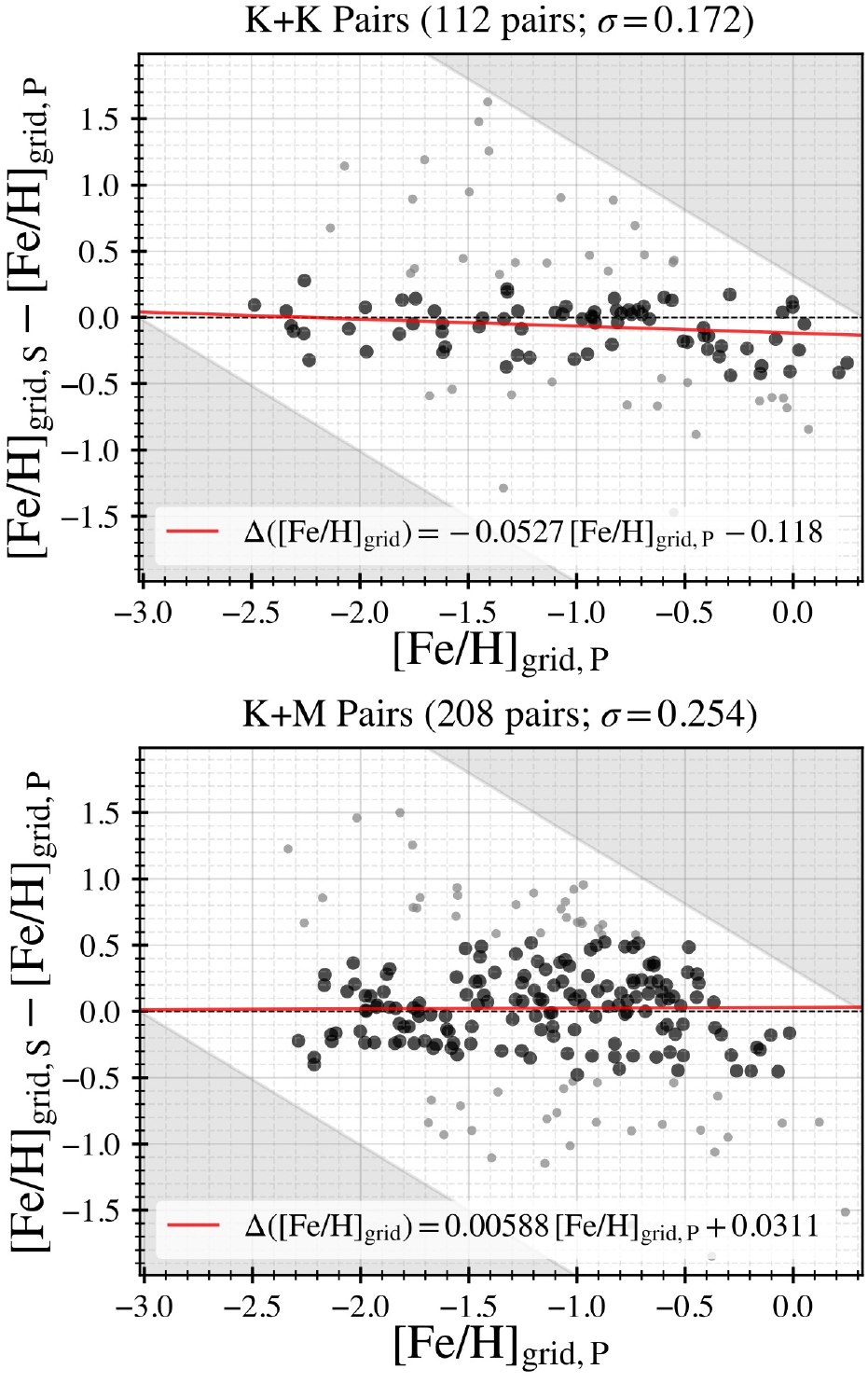}
\caption{Same as Figure~\ref{fig:fig9}, but using the metallicity estimates from the revised grid, after corrections to the spectroscopic metallicity estimates of the M dwarf calibrators. The revised grid brings the photometric metallicity values of the K+M pairs in agreement, which shows that the grid is self-consistent. \label{fig:fig11}}
\end{figure}

As we demonstrated in \S\ref{sec:binaries}, we compare metallicity estimates from the new grid of K+K pairs and K+M pairs to check the internal consistency of metallicity estimates from the grid. We use the same sample shown in Figure~\ref{fig:fig9}, but exclude stars with $\mathrm{[Fe/H]_{grid}} = -3.0$ or $0.3$ dex, which removes two binary systems from K+M pairs initially featured in Figure~\ref{fig:fig9}. Figure~\ref{fig:fig11} again shows the $\Delta\mathrm{[Fe/H]_{grid}}$ distribution as a function of $\mathrm{[Fe/H]_{grid, P}}$, but now displaying the results estimated from the revised grid. The red lines in both panels are the linear fits to the data with $-2.5 \le \mathrm{[Fe/H]_{grid}} \le 0.0$. Compared to Figure~\ref{fig:fig9}, the dispersion of K+M pairs is  about to be the same, however the systematic offset (red line; $\sim 0.031$ dex) is brought down very close to zero. 

\subsubsection{Validation and Caveats of the Photometric Metallicity Grid}\label{sec:caveats}

One way to validate the photometric metallicity grid is to compare photometric metallicities of stars in the collective dataset described in \S~\ref{sec:phot_feh} to spectroscopic metallicities. If our grid accurately estimated photometric metallicities, the difference between photometric and spectroscopic metallicities would be small.

\begin{figure*}
\centering
\includegraphics[width=\textwidth]{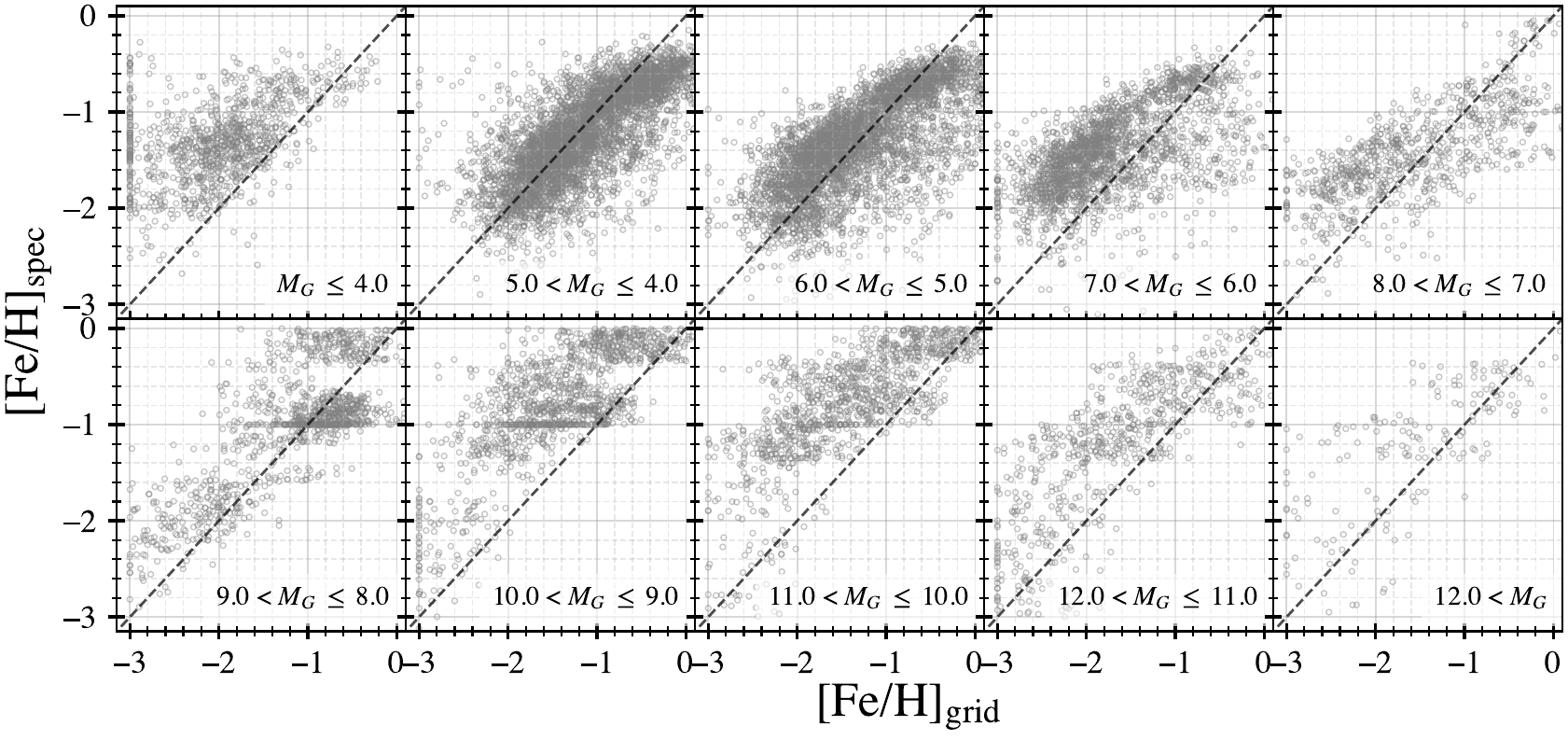}
\caption{Scatter plots displaying the distribution of stars in the collective dataset grouped by absolute $G$ magnitudes (increasing from top-left to bottom-right) in the metallicity space. Stars close to the main-sequence turn-off point ($M_G < 4.0$) and low-mass stars ($7.0 \le M_G < 12.0$) are showing a significant offset from the spectroscopic measurements.
\label{fig:fig12}}
\end{figure*}

Figure~\ref{fig:fig12} presents the difference between spectroscopic metallicities and the grid estimates of stars in the collective dataset, grouped by their absolute $G$ magnitudes. It appears that the metallicity difference is larger in early-type stars close to the main-sequence turn-off point ($M_G < 4.0$) and in low-mass stars including late K-type stars and all M dwarfs ($7.0 \le M_G < 12.0$).

The arbitrary offset we applied to adjust the grid for low-mass stars likely explains why low-mass stars show underestimated grid metallicities. Since the arbitrary offset we applied was introduced to bring in agreement the metallicity scales of K and M dwarfs, this suggests that the grid metallicity values are likely correct, and that the real problem lies with the spectroscopic metallicity estimates, which appear to overestimate the true metallicity of the M dwarfs.

For early-type stars, on the other hand, the reason they have significant offset is due to poor alignment of our grid on the bluer end of the main sequence as shown in the right panel in Figure~\ref{fig:fig10}. Due to the misalignment at the blue-end, our grid tends to overestimate metallicities for early-type dwarfs.

Another way to verify the applicability of the photometric metallicity grid for our entire catalogue is to verify if the distribution of stars consistently aligns with the grid for bins of different distances and Galactic latitudes. Since the local halo population is expected to be relatively homogeneous within our survey range ($d < 2$ kpc), subsets of stars should all consistently align with the grid.

\begin{figure*}
\centering
\includegraphics[width=\textwidth]{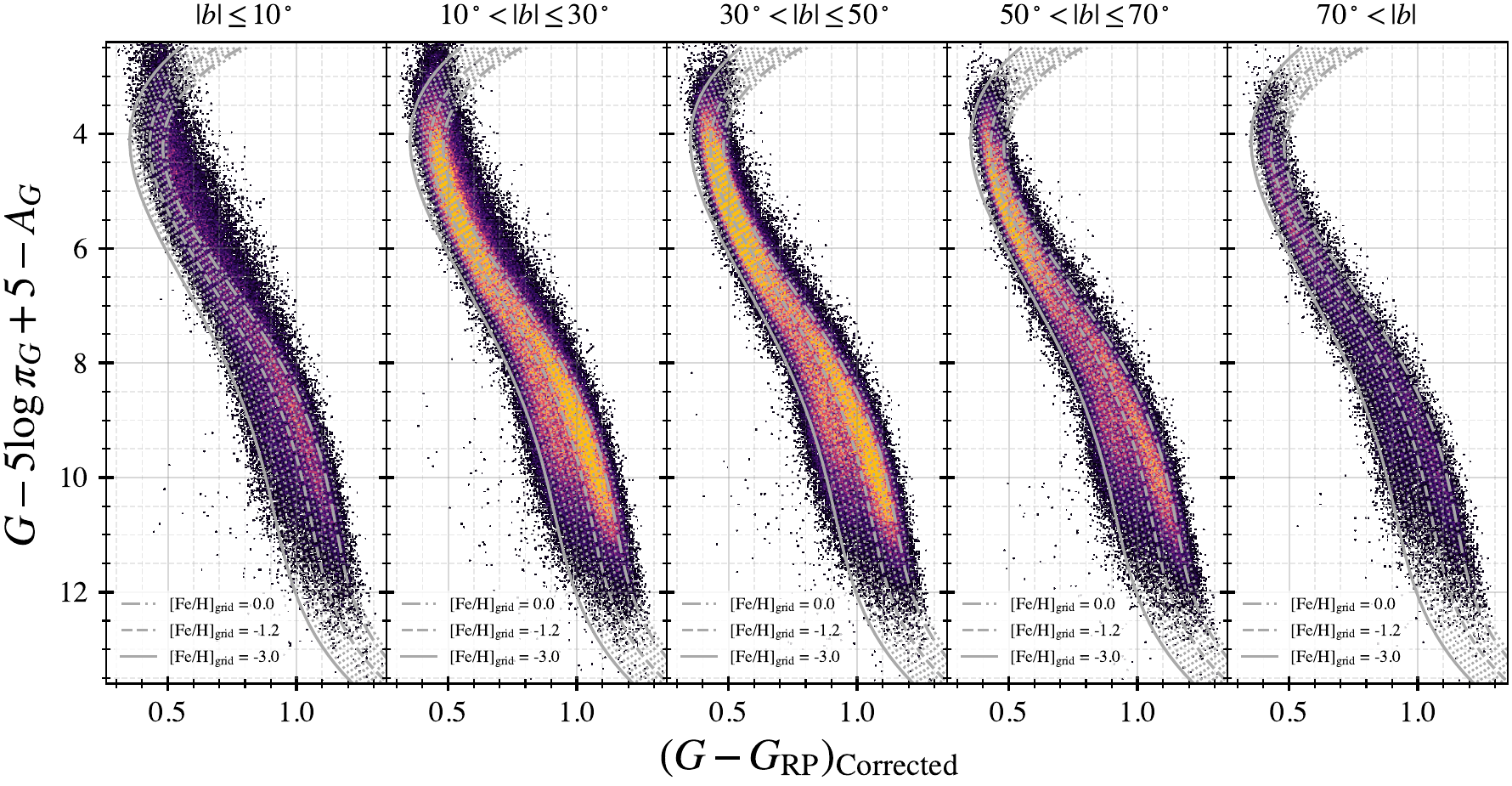}
\caption{CMDs of stars in our nearby halo catalogue, grouped in bins of Galactic latitude. The photometric metallicity grid lines with $\mathrm{[Fe/H]_{grid}} = 0.0, -1.2$, and $-3.0$ dex are shown in grey dash-dotted, dashed, and solid lines. The colour distribution of the late-type, low-mass population ($M_{G} \ge 8.0$) appears to be similar, regardless of Galactic latitude. Early-type stars in low Galactic latitude ($|b| \le 30^{\circ}$), on the other hand, show a noticeable drift to the red which suggests reddening effects, and likely results in an overestimate of their $\mathrm{[Fe/H]_{grid}}$. Low-mass stars are less susceptible to reddening errors because they are systematically closer, which makes their photometric metallicities more reliable.
\label{fig:fig13}}
\end{figure*}

Figure~\ref{fig:fig13} presents the grid overplotted on the entire distribution of halo candidates, for five subgroups based on their Galactic latitudes. The grid aligns well with the distribution of red, faint stars, in each one of the bins, which means the mean distribution of colours is independent of Galactic latitude. This suggests that low-mass stars ($M_G \ge 8$) in the catalogue remain largely unaffected by reddening biases, and that their photometric metallicity estimates are relatively reliable, regardless of position on the sky. This is not the same for stars of higher mass ($M_G < 8$). It is clear from Figure~\ref{fig:fig13} that earlier-type stars are found to be systematically redder than the grid at lower Galactic latitudes. While this could in part be explained with a higher fraction of unresolved systems at low-galactic latitudes, the most likely explanation is that these stars suffer from reddening effects, despite our earlier attempts to correct for this (see \S \ref{sec:rpmdiagram}). Considering this, $\mathrm{[Fe/H]_{grid}}$ of early-type stars in the low Galactic latitude may be overestimated. 

Aside from this general colour drift with Galactic latitude, we see that the early-type stars in our sample, notably stars in the main-sequence turnoff ($M_G \sim 3.5$) do not extend to the lowest metallicity grid line $\mathrm{[Fe/H]} \sim -3.0$, while later type stars do in significant numbers. This suggests that the envelope of the most metal-poor objects is not defined in a consistent way. In consequence, the bright end of the grid ($M_G < 5.5$) seems to be systematically shifted to blue by $0.014$ in $\grp$, which likely causes that $\mathrm{[Fe/H]}_{\mathrm{grid}}$ are overestimated by $0.3$ dex compared with lower-mass objects.

\begin{figure*}
\centering
\includegraphics[width=\textwidth]{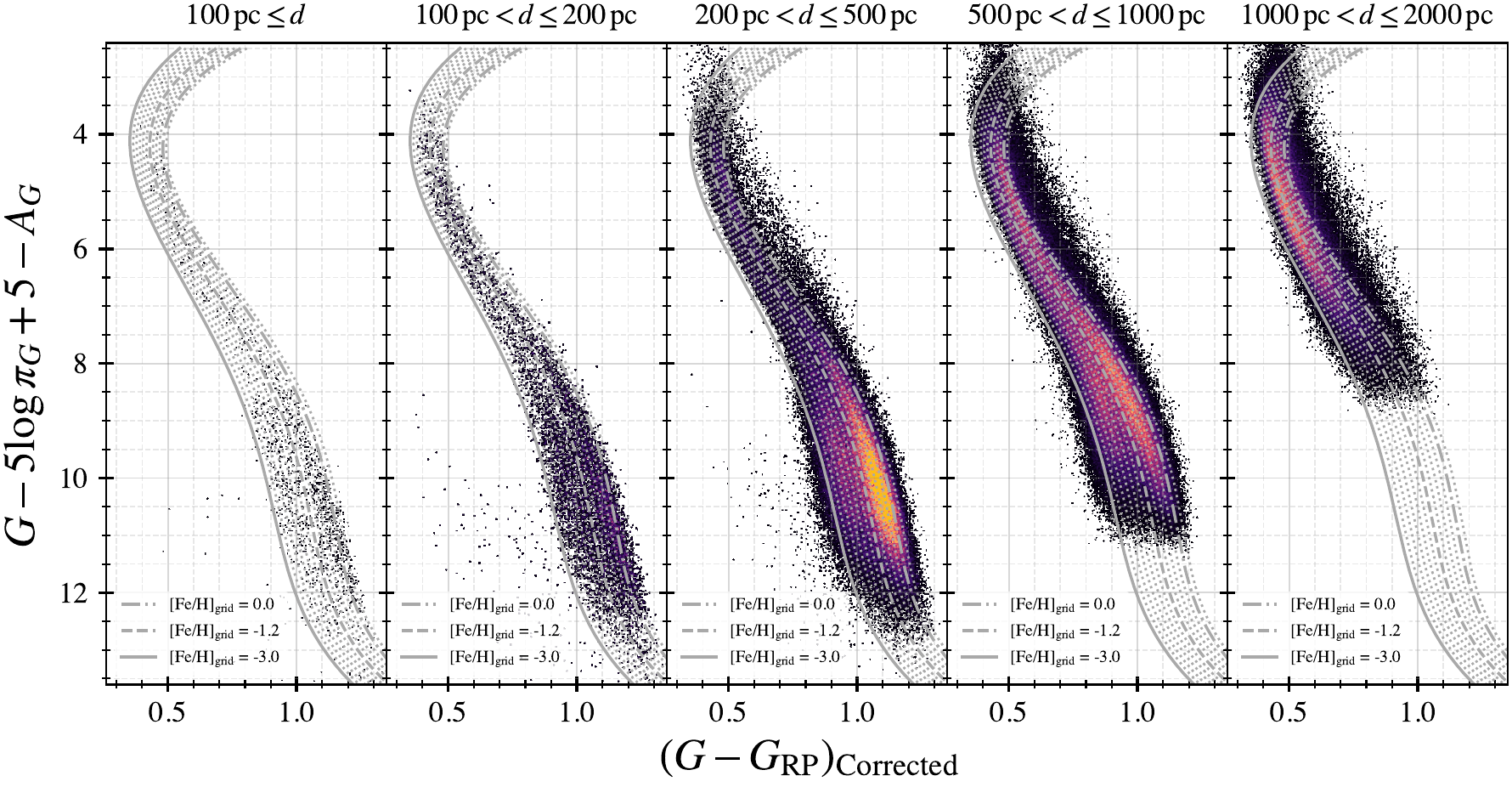}
\caption{CMDs of stars grouped in accordance with distance. The grid lines with $\mathrm{[Fe/H]_{grid}} = 0.0, -1.2$, and $-3.0$ dex are shown in grey dash-dotted, dashed, and solid lines. Although we do not limit distance or magnitude in our sample, the volume of the catalogue is limited within 2 kpc from the Sun due to the faint brightness of main-sequence stars. More than half of stars in the catalogue are late-type dwarfs ($M_G > 7.5$) that are mostly found in a volume of $500$ pc.
\label{fig:fig14}}
\end{figure*}

Figure~\ref{fig:fig14} shows the distribution of local halo stars, this time in bins of distance. This is a good illustration of the dominance of low-mass stars in the halo population, as late-type objects clearly dominate the bins with distances $d < 500$ pc. More than half of our halo candidates are late-type dwarfs ($M_G > 7.5;\,\,321,879$ stars), which are found within 1 kpc, as can be seen in Figure~\ref{fig:fig14}. Most early-type dwarfs ($M_G \le 7.5;\,\,229,335$ stars), on the other hand, are found at larger distances, which expands the volume of our catalogue to $d \sim 2$ kpc. Even so, the volume completeness of the catalogue is limited to $d \sim 500$ pc, considering the main source of our catalogue is mostly late-type dwarfs. Figure~\ref{fig:fig14} also shows that our catalogue becomes magnitude-limited beyond $d > 500$ pc, and suffers from significant incompleteness at the low-mass end. The catalogue is however statistically complete at least for stars with $M_G < 12.0$ to a distance $d < 500$ pc.

\subsection{Comparison to Photometric Metallicities in SkyMapper DR2}

\citet[hereafter C21]{chiti:21} have recently reported photometric metallicities ($\mathrm{[Fe/H]_{SM2}}$) for $\sim 720{\mathrm ,}000$ stars with $-3.75 \lessapprox \mathrm{[Fe/H]_{SM2}} \lessapprox -0.75$. Their photometric metallicities were derived by using SkyMapper $u$, $g$, $i$ photometry that is sensitive to stellar metallicities and the dependency of SkyMapper $v$ flux on the Ca II K absorption features; these metallicity estimates are thus entirely colour-based. To compare our result to C21, we crossmatched our catalogue to C21 and recovered $3271$ stars with good photometric metallicity estimates (less than 10\% of fractional $\mathrm{[Fe/H]_{SM2}}$ errors) from SkyMapper.

\begin{figure*}
    \centering
    \includegraphics[width=\textwidth]{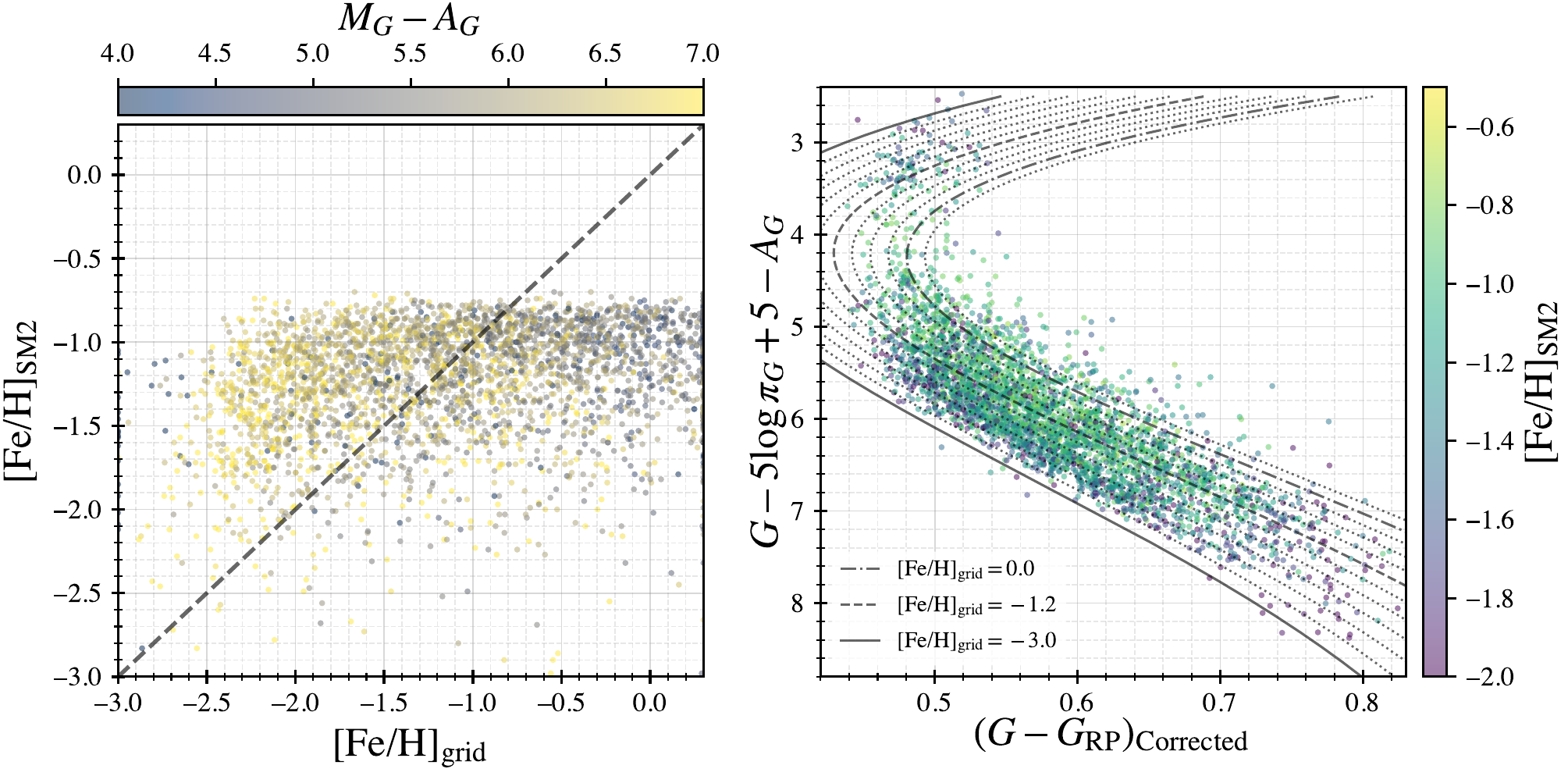}
    \caption{Left: Scatter plot, colour-coded by absolute $G$ magnitude, of the distribution of stars in both C21 and our catalogues in the metallicity space. Right: CMD of stars in both catalogues, colour-coded by photometric metallicities from \citet{chiti:21}. Each grid line shows our photometric metallicity grid in 0.3 dex increment.
    \label{fig:fig15}}
\end{figure*}

In Figure~\ref{fig:fig15}, the left panel compares the SkyMapper metallicities to our grid metallicity estimates, with the stars colour-coded by their absolute $G$ magnitudes. Due to the metallicity range of C21 ($-3.75 \lessapprox \mathrm{[Fe/H]_{SM2}} \lessapprox -0.75$), a selection effect is clearly shown in around $\mathrm{[Fe/H]_{SM2}} \sim -0.75$. Even Considering this, however, our grid metallicity estimates have very poor agreement with C21's values. Stars showing the largest metallicity differences tend to be ones at the bright-end with solar metallicities from the grid (darker coloured points) or those at the faint-end with very low grid metallicities. A right panel in Figure~\ref{fig:fig15} shows that these stars are mostly early-type stars, evolved stars, late K-dwarfs, early M-dwarfs, and/or binaries.

\subsection{Luminosity Function for Stars in the Local Halo}\label{sec:lf}
If our catalogue is statistically complete within $500$ pc from the Sun, it is possible to generate a luminosity function (LF) of the local halo by simply counting stars. As the observed LF is biased by environmental factors, we took into account all systematics that may affect an accurate count of the stars, securing a sufficient number of stars to conduct a statistical study, completeness of a volume of the study, precise distances from Gaia parallaxes, extinction correction, identification of unresolved binaries in the CMD, and stellar metallicity \citep{bochanski:10}. 

\begin{figure*}
    \centering
    \includegraphics[width=\textwidth]{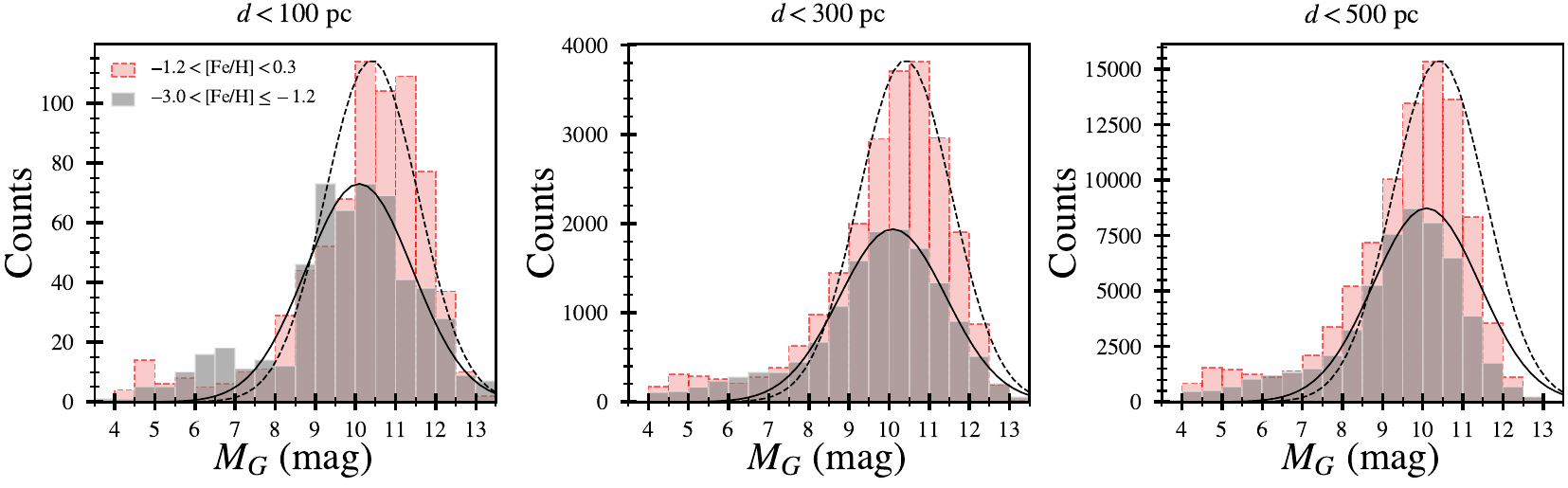}
    \caption{Histograms of stars in the Solar neighbourhood, grouped in bins of increasing volume. Red histograms represent the number counts of metal-rich stars (the red sequence; $-1.2 < \mathrm{[Fe/H]} < 0.3$), and grey histograms show the counts of metal-poor stars (the blue sequence; $-3.0 < \mathrm{[Fe/H]} \le -1.2$), both as a function of absolute magnitude $M_G$. Black dashed and solid lines in each panel are the Gaussian fits to the histograms of stars within $300$ pc from the Sun, but rescaled in the other two plots to match the amplitude. 
    \label{fig:fig16}}
\end{figure*}

Figure~\ref{fig:fig16} shows histograms of absolute magnitude $M_G$ for local halo in increasing volumes for stars within $100$ pc (left), $300$ pc (centre), and $500$ pc (right), respectively. We first removed evolved stars at the bright end of the main sequence based on this empirical cut: $(M_G)_{\mathrm{evolved}} < 4.264(\grp) + 2.188$. We then divided all remaining stars into two groups, selecting stars in the red, metal-rich sequence (red histogram, $-1.2 < \mathrm{[Fe/H]_{\mathrm{grid}}} < 0.3$) and the blue, metal-poor sequence (grey histogram, $-3.0 < \mathrm{[Fe/H]_{\mathrm{grid}}} \le -1.2$) in the CMD. Outliers (such as some unresolved binaries  or high-reddening stars) that are systematically bluer/redder than the grid are excluded as we only select stars falling on the specified range of the metallicity grid. Black dashed and solid lines in each panel are Gaussian fits to the histograms of low-mass stars ($M_G > 7.5$) within $300$ pc from the Sun. These Gaussian fits are rescaled for amplitude and plotted in the $d < 100$ pc and $d < 500$ pc distribution, to help compare the three distributions. Table~\ref{tab:tab4} presents star counts for stars in each metallicity group. We assume that the distribution of absolute magnitude, $M_G$, has a Poisson distribution to estimate errors.

It is notably clear that low-mass stars are indeed dominant. Assuming we define ``intermediate-mass stars" as AFG dwarfs, defined to be stars with $M_G \le 7.5 $, and ``low-mass stars" as KM dwarfs, defined to be stars with $M_G > 7.5$, then we can calculate $N_{\mathrm{AFG}}$ as the total number of intermediate-mass stars in each distribution, and $N_{\mathrm{KM}}$ as the total number of low-mass stars in each distribution. The number ratios of low-mass stars $N_{\mathrm{KM}}/N_{\mathrm {tot}}$ in both populations are more than $86$\% (see Table~\ref{tab:tab4}). The metal-rich population, however, appears to have a higher number ratio compared to the metal-poor population, $\sim 92$\% for metal-rich stars to $\sim 88$\% for metal poor stars. One caveat is that some unresolved binaries in the blue (metal-poor) sequence may be shifted into the red (metal-rich) sequence due to those stars being overluminous at a given colour, and this could bias the ratios depending on what the multiplicity fraction is. By comparison, the general statistics for field (Galactic disk) stars in the Solar neighbourhood shows that the fraction ratio of low-mass stars is only about $80$\%\footnote{This value is empirically obtained from the full Gaia EDR3 subset of nearby stars within $100$ pc from the Sun. We applied simple selection criteria to select ``clean" sample, which is: $\sigma_\pi/\pi < 0.10$, flux errors in BP/RP filters better than $10$\%, and \texttt{RUWE} $< 1.4$. After removing red giants, white dwarfs, and the overluminous stars above the main sequence, we obtained $\sim 81$\% as the ratio of the low-mass stars to total number of main-sequence stars within $100$ pc from the Sun.}, each group still contains about $7$\% to $12$\% of unresolved binaries as contaminants.

Interestingly, the Gaussian fits (black dashed and solid lines in Figure~\ref{fig:fig16}) for stars within $300$ pc are well aligned with the distributions of stars within $100$ pc, which indicates that there is no luminosity bias at least to a distance range $d \sim 300$ pc. The faint-end of the Gaussian fits for stars in both metallicity groups, however, becomes slightly depleted in the $d \sim 500$ pc volume. The overall low-mass star ratio also decreases as the volume expands (see Table~\ref{tab:tab4}). These observations indicate that survey incompleteness starts presenting around $500$ pc. 

Each metallicity group shows a slightly different bimodal distribution with peak absolute magnitudes of $M_G \sim 5.0$ and $M_{G} \sim 10.5$ for the metal-rich stars and $M_{G} \sim 7$ and $M_G \sim 10$ for the metal-poor stars. The difference between the bright magnitude peaks can be explained by the fact that intermediate-mass (AFG) stars in the halo may have formed and evolved earlier than those in the thick disk did. If most intermediate-mass stars are now in the form of evolved giants or white dwarfs, the distribution of the remaining main-sequence objects will show a deficit at the intermediate-mass end. Intermediate-mass stars in the metal-rich group, on the other hand, has a peak absolute magnitude at $M_{G} \sim 5$, which is a close fit to the Sun's absolute magnitude in Gaia $G$ band \citep[$M_{G, \odot} = 4.68$ mag;][]{andrae:18}, and suggests that Sun-like stars remain largely unevolved in that group. 

The difference in the low-luminosity peaks is most likely caused by metallicity effects, due to the fact that the mass-luminosity relation (MLR) varies with metallicity. A metal-poor object is shown to be more luminous and bluer compared to its solar-metallicity counterpart with the same mass \citep{sandage:59, bochanski:10}. It is thus possible that the {\it mass function} peaks at the same value for both groups, but the {\it luminosity function} peaks at different values because the {\it mass-luminosity} relationship is different for the metal-rich and metal-poor groups.

The other noticeable difference between the two groups is the number counts for the low-mass stars (see Table~\ref{tab:tab4}); The number ratios of the low-mass stars in the metal-poor group to those in the metal-rich group ($N_{\mathrm{KM, MP}}/N_{\mathrm{KM, MR}}$) is $0.72 \pm 0.044$ ($d < 100$ pc), $0.58 \pm 0.0065$ ($d < 300$ pc), and $0.59 \pm 0.0034$ ($d < 500$ pc), respectively. This may be interpreted as the signature of a more top-heavy initial mass function (IMF) in the star-forming period in the early Universe. A top-heavy IMF is an initial mass distribution that contains more massive stars than the canonical mass function \citep[i.e.,][]{kroupa:01}. The idea of a top-heavy IMF in the early Universe originates in the cooling behaviour of the gas that depends on its metal components. As cooling in low-metallicity environments is less efficient, the early star-forming environment, where the metal components were much fewer than present, is most likely to have favoured massive and intermediate mass stars over low-mass stars \citep{marks:12}. If the local halo population has this top-heavy IMF in the early Universe and if the most massive stars have evolved off to become giants and white dwarfs, it is explicable why the present-day luminosity function has the form seen in Figure~\ref{fig:fig16}.

\begin{table*}
\caption{Star Counts for stars in the Local Halo.}
\label{tab:tab4}
\resizebox{\textwidth}{!}{%
\begin{tabular}{c|ccc|ccc|ccc}
\hline
\multirow{2}{*}{Group} & 
\multicolumn{3}{c|}{Counts for intermediate-mass Stars ($N_{\mathrm{AFG}}$)} & 
\multicolumn{3}{c|}{Counts for Low-mass Stars ($N_{\mathrm{KM}}$)} & 
\multicolumn{3}{c}{$N_{\mathrm{KM}}/N_{\mathrm{tot}}$} \\  \cline{2-10}
                       & 
$100$ pc & $300$ pc & $500$ pc & 
$100$ pc & $300$ pc & $500$ pc & 
$100$ pc & $300$ pc & $500$ pc \\
\hline
\hline
Metal-rich & 
\multirow{2}{*}{$56 \pm 7$} & \multirow{2}{*}{$1\mathrm{,}994 \pm 45$} & \multirow{2}{*}{$10\mathrm{,}288 \pm 101$} &
\multirow{2}{*}{$657 \pm 26$} & \multirow{2}{*}{$21\mathrm{,}477 \pm 147$} & \multirow{2}{*}{$81\mathrm{,}562 \pm 286$} &
\multirow{2}{*}{$0.921 \pm 0.0360$} & \multirow{2}{*}{$0.915 \pm 0.00624$} & \multirow{2}{*}{$0.888 \pm 0.00311$} \\
($-1.2 < \mathrm{[Fe/H]_{\mathrm{grid}}} < 0.3$) & & & & & &  &  & & \\
\\
Metal-poor & 
\multirow{2}{*}{$68 \pm 8$} & \multirow{2}{*}{$1\mathrm{,}684 \pm 41$} & \multirow{2}{*}{$7\mathrm{,}393 \pm 86$} &
\multirow{2}{*}{$475 \pm 22$} & \multirow{2}{*}{$12\mathrm{,}357 \pm 111$} & \multirow{2}{*}{$48\mathrm{,}071 \pm 219$} & 
\multirow{2}{*}{$0.875 \pm 0.0401$} & \multirow{2}{*}{$0.880 \pm 0.00792$} & \multirow{2}{*}{$0.867 \pm 0.00395$} \\
($-3.0 < \mathrm{[Fe/H]_{\mathrm{grid}}} \le -1.2$) & & & & & &  &  & & \\
\hline
\end{tabular}%
}
\end{table*}

From the observed luminosity function, it would technically be possible to estimate an IMF of the local halo only if we knew the MLR for main-sequence stars in the halo. Since the IMF is the number distribution of masses of stars that have been just born, knowing the IMF translates into how many stars formed in a given mass interval in the birth environment \citep{kroupa:93, covey:08}. The MLR at the low-mass end remains difficult to calibrate because of the difficulty in observing a sufficient number of low-mass dwarfs in binary systems, for which the individual masses of the stars can be independently derived \citep{henry:99, bochanski:10, benedict:16}. The situation is especially dire for metal-poor stars, to the rarity of know low-mass binary systems in the halo population. To calibrate the MLR at the low-mass end for thick disk and halo stars would require identifying and monitoring a significant number of low-mass eclipsing systems from these metal-poor population. \citet{jao:16} made a start on building an empirical MLR for the low-mass, metal-poor stars using 10 subdwarfs in five double-line spectroscopic binaries. They demonstrated several challenges that they had to face while building the MLR, and one of the reason was the rarity of metal-poor calibrators in the Solar neighbourhood. Today, the fastest way to search for these MLR calibrators (i.e., eclipsing binaries) would be to search for them among light curves of halo stars that might have been observed by the TESS/Kepler missions. Our present catalogue of low-mass, metal-poor stars identifies the best candidates (nearest and brightest) for carrying out this program.

\subsection{Comparison to the \texttt{PARSEC} Isochrones}

\begin{figure}
\centering
\includegraphics[width=\columnwidth]{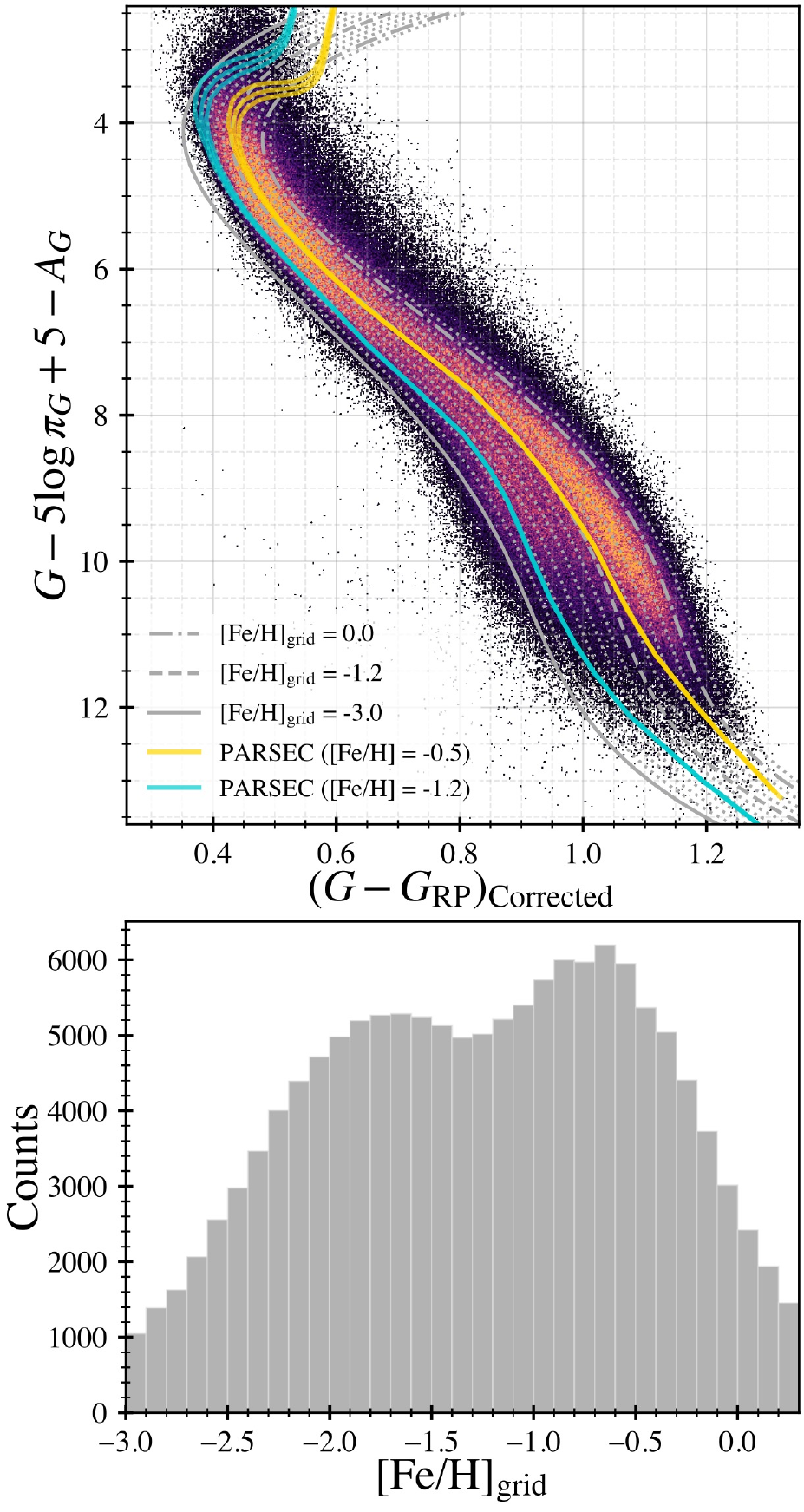}
\caption{Top: CMD of all halo candidates in the catalogue, with the empirical photometric metallicity grid overplotted (grey lines). The very sparse objects found at the lower-left of the main-sequence are most likely to be WD+M pairs, which may belong to the Galactic disk, not the halo. The yellow and blue lines represent the \texttt{PARSEC} theoretical model isochrones \citep{bressan:12} with $\mathrm{[Fe/H]} = -0.5$ and $-1.2$ and ages of $10$, $11$, and $12$ Gyr from the left to right. Bottom: histogram of $\mathrm{[Fe/H]_{grid}}$ of halo dwarf candidates in the high Galactic latitude ($|b| > 50^{\circ}$). The distribution shows bimodality consistent with two populations with $\mathrm{[Fe/H]_{grid}} \simeq -0.7$ and $-1.7$, which reflects the dual main sequences shown in the top panel.
\label{fig:fig17}}
\end{figure}

We present a distribution of all halo candidates in the CMD in the top panel in Figure~\ref{fig:fig17}. Grey lines in the CMD are the grid lines with $\mathrm{[Fe/H]_{grid}}$ from $-3.0$ to $0.3$ in $0.3$ dex increments. As recent studies reported \citep{gaia:18a, gaia:21}, our sample shows dual main sequences in the CMD. The upper and lower main sequences are aligned with the grid lines with $\mathrm{[Fe/H]_{grid}} = -0.7$ and $-1.7$. As we described in \S~\ref{sec:caveats}, we do recognise that the overall grid is shifted to blue by $\grp = 0.014$ in the range of $M_G < 5.5$, which causes misalignment of the grid to actual distribution of stars. The bottom panel in Figure~\ref{fig:fig17} shows a histogram of $\mathrm{[Fe/H]_{grid}}$ for stars in high Galactic latitude ($|b| > 50^{\circ}$) with metallicities of $-3.0 < \mathrm{[Fe/H]_{grid}} < 0.3$. We remove main sequence turn-off stars using the empirical colour-magnitude cut: $M_{G} > 4.264\times\grp + 2.188$. The distribution of the grid metallicity estimates reflects the dual main sequence with peak metallicity values of $\mathrm{[Fe/H]_{grid}} = -0.7$ and $-1.7$.

A grid line with $\mathrm{[Fe/H]_{grid}} = -1.2$, shown in a dashed line, matches the gap between the upper and lower main sequences. The grid line also matches the gap between the dual peaks in the histogram shown in the bottom panel in Figure~\ref{fig:fig17}. Considering the result reported from \citep{gaia:21}, our value is significantly lower than their parameters for the \texttt{PARSEC} isochrone with $\mathrm{[Fe/H]} = -0.5$ and age of $11$ Gyr, although it may not be appropriate to directly compare their result to ours as we do not have age information for our stars. To compare the \texttt{PARSEC} isochrones to our grid, we plot the isochrones with $\mathrm{[Fe/H]} = -0.5$ (yellow lines) and $-1.2$ (blue lines) in the CMD. To see the age effect, we plot three isochrones with the same metallicity, but different ages of $10$, $11$, and $12$ Gyr from left to right. Our calibrated grid line with $\mathrm{[Fe/H]_{grid}} = -1.2$ is most similar to the isochrone with $\mathrm{[Fe/H]} = -0.5$ and age of $11$ Gyr. Our grid is an empirical fit to the data based on available spectroscopic metallicity estimates; it remains unclear why our grid value is $0.7$ dex less than the estimate from the model isochrones, but this points to at least inconsistencies between the predicted colours from evolutionary models and spectroscopic metallicities derived from stars of the same colour and absolute magnitude.

\section{A catalogue of main-sequence stars in the local Galactic thick disk and halo in the Gaia EDR3}\label{sec:catalogue}
We present a final catalogue of $551,214$ main-sequence stars in the local Galactic thick disk and halo in the Gaia EDR3. Our catalogue provides basic information in Gaia EDR3, including source id, position (R.A., Dec.), kinematic parameters (parallax, proper motion in R.A. and Dec.) and their uncertainties, photometric magnitudes and colours ($G$ and $\grp$ colour), and and the adopted extinction value ($A_V$) based on an application of the 3D reddening map of \citet{green:19}. Finally, the catalogue lists our photometric metallicity estimates from both the KNN regressor and from the calibrated colour-magnitude-metallicity grid. Table~\ref{tab:tab4} presents description of columns in the catalogue and Table~\ref{tab:tab6} displays a portion of the catalogue with 10 sample stars. The full catalogue is provided electronically in a machine-readable format.

\begin{table}
\caption{Columns in the catalogue}
\centering
\label{tab:tab5}
\resizebox{\columnwidth}{!}{%
\begin{tabular}{cl}
\hline
Header & Description \\
\hline
\hline
 \texttt{source\_id} & {\it Gaia} EDR3 unique source identifier \\
 R.A. & Right ascension in J2016.0 (deg) \\
 Dec. & Declination in J2016.0 (deg)  \\
 $\pi$ & Parallax (mas) \\
 $\sigma_{\pi}$ & Standard error of parallax (mas) \\
 $\mu_{\alpha}$ & Proper motion in right ascension direction (mas yr$^{-1}$) \\
 $\mu_{\delta}$ & Proper motion in declination direction (mas yr$^{-1}$) \\
 $G$ & $G$-band mean magnitude (mag) \\
 $\grp$ & $\grp$ colour (mag) \\
 $E(\grp)$ & $\grp$ colour excess (mag) \\
 $A_{G}$ & {\it Gaia} $G$-band extinction (mag) \\
 $\mathrm{[Fe/H]_{KNN}}$ & Metallicity estimate from the KNN regressor (dex) \\
 $\mathrm{[Fe/H]_{grid}}$ & Metallicity estimate from the photometric metallicity grid (dex) \\
\hline
\end{tabular}%
}
\end{table}

\begin{table*}
\caption{The Catalogue of Candidates in the Galactic Thick Disk and} Halo in {\it Gaia} EDR3
\label{tab:tab6}
\resizebox{\textwidth}{!}{%
\begin{tabular}{lcccccccccccc}
\hline
\texttt{source\_id} & R.A. & Dec. & $\pi$ & $\sigma_{\pi}$ & $\mu_{\alpha}$ & 
 $\mu_{\delta}$ & $G$ & $\grp$ & $E(\grp)$ & $A_{G}$ & $\mathrm{[Fe/H]_{KNN}}$ & $\mathrm{[Fe/H]_{grid}}$ \\
 & (deg) & (deg) & (mas) & (mas) & ($\masyr$) & ($\masyr$) & (mag) & (mag) & & (mag) & (dex) & (dex)\\
\hline
\hline
4489782790598521344 & $268.10093$ & $ 11.066672$ &	$5.2962$ &	$0.0926$ &    $-70.637$ &	$-226.94$ &	$17.316$ &	$1.146$ &	$0.000$ &	$0.000$ &	$-0.218$ &	$0.0144$ \\
4490275509244724608 & $262.78433$ & $ 8.3478274$ &	$5.8721$ &	$0.0609$ &    $-157.28$ &	$-116.45$ &	$16.624$ &	$1.125$ &	$0.000$ &	$0.000$ &	$-0.0908$ &	$-0.0104$ \\
4489082161172907136 & $265.86335$ & $ 9.2758875$ &	$4.0087$ &	$0.0949$ &    $-11.188$ &	$-109.42$ &	$17.443$ &	$1.184$ &	$0.091$ &	$0.371$ &	$-0.149$ &	$0.0741$ \\
4490338005313126912 & $262.41825$ & $ 8.5301606$ &	$2.4762$ &	$0.153 $ &    $-47.539$ &	$-49.179$ &	$18.316$ &	$1.140$ &	$0.067$ &	$0.274$ &	$-0.188$ &	$-0.198$ \\
4489611640444691968 & $267.01876$ & $ 10.210906$ &	$3.4534$ &	$0.142 $ &    $-78.226$ &	$-68.369$ &	$17.959$ &	$1.164$ &	$0.073$ &	$0.299$ &	$-0.178$ &	$-0.179 $ \\
4488862155769809536 & $268.83964$ & $ 10.024653$ &	$5.6977$ &	$0.0358$ &    $-79.302$ &	$-229.72$ &	$15.453$ &	$1.065$ &	$0.000$ &	$0.000$ &	$-0.168$ &	$0.0902$ \\
4490471492895500544 & $260.42788$ & $ 8.4753347$ &	$3.3793$ &	$0.156 $ &    $-46.741$ &	$-74.090$ &	$18.245$ &	$1.214$ &	$0.097$ &	$0.399$ &	$-0.205$ &	$0.125$ \\
6255613166476392320 & $227.78030$ & $-21.167114$ &	$2.7782$ &	$0.148 $ &    $-59.113$ &	$-29.145$ &	$18.180$ &	$1.130$ &	$0.073$ &	$0.299$ &	$-0.231$ &	$-0.435$ \\
6252911052232066688 & $231.01264$ & $-20.714646$ &	$2.6004$ &	$0.0180$ &    $-112.19$ &	$ 8.5257$ &	$12.358$ &	$0.564$ &	$0.085$ &	$0.349$ &	$-0.583$ &	$-0.486$ \\
6256890008712715520 & $226.89827$ & $-19.280277$ &	$2.8704$ &	$0.182 $ &    $-7.5345$ &	$-81.105$ &	$18.361$ &	$1.164$ &	$0.055$ &	$0.225$ &	$-0.179$ &	$-0.0553$ \\
... & ... & ... & ... & ... & ... & ... & ... & ... & ... & ... & ... & ... \\
\hline
\multicolumn{13}{l}{$^*$\footnotesize{A full machine-readable catalogue is available online.}} \\
\end{tabular}}
\end{table*}

\subsection{Confirmation of Halo Kinematics}
The halo population is kinematically distinctive from the disk population, most notably displaying a very large ``asymmetric drift" relative to the local standard of rest. We verify the halo status of the stars in our catalogue by checking the global kinematics of the stars on our catalogue. Conducting kinematic analysis normally requires radial velocities to obtain their full (3D) spatial motions and constrain their Galactic orbit. However, a large majority of stars in our catalogue do not currently have radial velocity measurements and may not be available in the future Gaia data sets due to their faint brightness \citep{sartoretti:18}. Hence, we use an indirect approach to have a glimpse of their global kinematics by using tangential velocities of stars in selective parts of the sky.

We follow the procedure described in \S 2.2 in \citet{kim:20}, which demonstrates how to use tangential velocities of stars in selected areas as projected $UVW$ velocities by rotating their Galactic coordinates. Since this method can provide its best result only if we apply it to stars in selective patches of the sky, we first select stars in the patches near both Galactic poles ($|b| > 70^{\circ}$; $36,801$ stars) and near the Galactic centre ($(l, b) = (0^{\circ},\, 0^{\circ})$; $11,400$ stars) and anticentre ($(l, b) = (180^{\circ},\,0^{\circ})$; $7,026$ stars) with a search radius of $20^{\circ}$. Tangential velocities of stars near the Galactic poles are parallel to Galactic $U$ and $V$ velocity vectors when applying the rotation vector to tilt their position and proper motion vectors by $90^{\circ}$ from the Galactic coordinate system. Tangential velocities of stars near Galactic centre and anticentre, on the other hand, can be directly projected to the Galactic $V$ and $W$ velocity except folding signs of $v_{t, l}$ of stars in $90^{\circ} \le l < 270^{\circ}$ to align them with stars in other patches. We should mention that our result has not been corrected by the Sun's motion with respect to the Galactic centre (Sgr A$^{*}$) and the motion of the local standard of rest.

\begin{figure*}
\centering
\includegraphics[width=\textwidth]{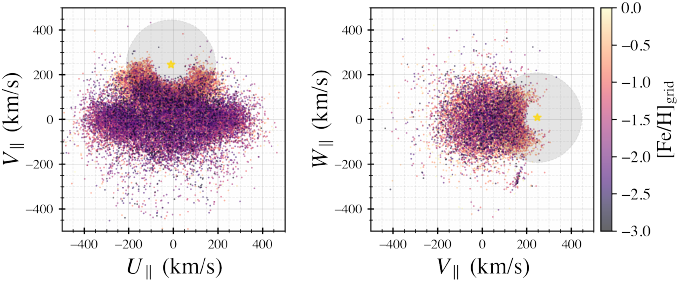}
\caption{Distribution of stars in our catalogue with colour-coded $\mathrm{[Fe/H]_{grid}}$ values in the projected Galactic $UVW$ spatial velocity space. A yellow star represents the Sun. Grey shaded area displays kinematic limits of the disk population ($V < 200$ km s$^{-1}$). Although these are not drawn by the actual $UVW$ velocities, a well-known stellar stream in the local halo, i.e., Gaia-Enceladus stream, is prominently shown in the $UV$ space (left), which seems to be dominant by the metal-poor population in our catalogue.
\label{fig:fig18}}
\end{figure*}

Figure~\ref{fig:fig18} shows the distribution of stars in Galactic $UVW$ velocity space, with their photometric metallicities represented by the colour-scale of the data points. The left panel shows stars in the patches near both Galactic poles in $U_{\parallel}$ and $V_{\parallel}$ velocity space, and the right panel shows all stars in the patches near Galactic centre and anticentre in $V_{\parallel}$ and $W_{\parallel}$ space. A star-shaped point represents the Sun's motion in each plane of projection. Shaded area shows the assumed kinematic limit of the disk population ($V < 200$ km s$^{-1}$). More than $64$\% of stars ($23,666$ stars) in the first group and $54$\% of stars ($10,038$ stars) in the second group are found outside the shaded area, which unambiguously places them with the local halo population. Stars inside this limit could technically represent the kinematic tail of the old disk or thick disk population. 

The distribution of stars clearly shows the large asymmetric drift expected of the local halo, and the most metal-poor objects notably show a distribution consistent with a mean motion $(U,\,V,\,W)=(0,\,0,\,0)$ in the Galactocentric frame, consistent with zero net angular momentum. The $UV$ distribution shows a clear metallicity gradient, with stars in the ``disk" region appearing to be significantly more metal-rich, while stars unambiguously in the halo looking more metal-poor. This supports the fact that the halo population is chemically more pristine than the disk population. The metallicity gradient is less evident in the $VW$ plane. However, the more metal poor stars do show a more spherical distribution and higher overall dispersion. The most prominent feature in the $UV$ plane is the elongated shape around $V_{\parallel} \approx 0$ km s$^{-1}$, which is the Gaia-Enceladus stream \citep{belokurov:18, myeong:18a, myeong:18b, haywood:18}. Most of stars in the stream appear in darker colour, which also indicates that they are relatively more metal-poor than the disk population that is located inside the shaded area. In the right panel, there is also a clump of metal-poor stars with $130 < V_{\parallel} < 190 $ km s$^{-1}$ and $-200 < W_{\parallel} < 310$ km s$^{-1}$, which may be kinematic members of the Helmi stream \citep{helmi:99, koppelman:18, koppelman:19a}. 

We will examine these kinematic features in more detail in a future paper, and only point them out qualitatively for now. The main takeaway for now is that our selection does recover much of the kinematic features expected of local halo stars. We also emphasise that the vast majority of our stars are low-mass K and M dwarfs, which means we are recovering the bulk of the local stellar halo.

\subsection{Comparison to \citet{koppelman:21}}
Since Gaia published their data for the first time in 2016, a number of studies based on Gaia have reported very exciting, new results on the kinematic makeup of the Galaxy. For example, \citet[hereafter KH21]{koppelman:21} have presented a catalogue of halo main-sequence stars in Gaia DR2 selected using an RPM diagram as we did in the present study. Since we share some similarity to their selection methods, we here compare our result to KH21.

KH21 identified a much larger set of $11,711,399$ stars as halo main-sequence stars using the RPM diagram. To identify the halo stars, they made a fit as a function of $\grp$ colour and $M_G$ to stars with precise parallaxes ($\sigma_\pi/\pi < 0.02$) and large tangential motions ($\vt > 300$ km s $^{-1}$) in the lower main sequence in the CMD, which was performed to minimise the contribution from the upper main sequence of stars with high-tangential-velocities. This fit was converted to a relationship between $\grp$ colour and tangential velocity so that they could use it to identify halo main-sequence stars in the range $200\,\, \mathrm{km\,s}^{-1} < \vt < 800$ km s$^{-1}$ in the RPM diagram. They obtained photometric distances of main-sequence stars from a colour-luminosity relationship, because only $\sim 10$\% of stars in the Gaia DR2 have most precise Gaia parallaxes ($\sigma_{\pi}/\pi < 0.2$) whereas having precise distances to distant stars is critical when analysing the kinematics of the Galaxy. Within their sample, $7,117,555$ stars have reliable photometric distances in the range $0.45 < \grp < 0.175$, which excludes main-sequence turn-off stars and low-mass stars. Assuming the radial velocities of stars to be zero, as radial velocity information is insufficient, they provided a kinematic analysis of these halo stars and found $\sim 83$\% of the stars are located near the vicinity of the Sun ($R < 8.2$ kpc). Their result recovered the local halo structures, such as the Gaia-Enceladus, the Helmi streams, and some other substructures including Sequoia and Thamnos.

The most significant difference between our catalogue and KH21 is the volume surveyed by the selection, which explains the difference in numbers. With our focus on the local halo populations, we selected from the subset of Gaia high-proper-motion ($\mu > 40$ mas yr$^{-1}$) stars with the parallax quality cut ($\sigma_\pi/\pi < 0.15$). Since this tends to select for relatively nearby sources, it is not surprising that the number of stars in KH21 is about 21 times greater than the number of stars in our catalogue. 

\begin{figure*}
\centering
\includegraphics[scale = 1.0]{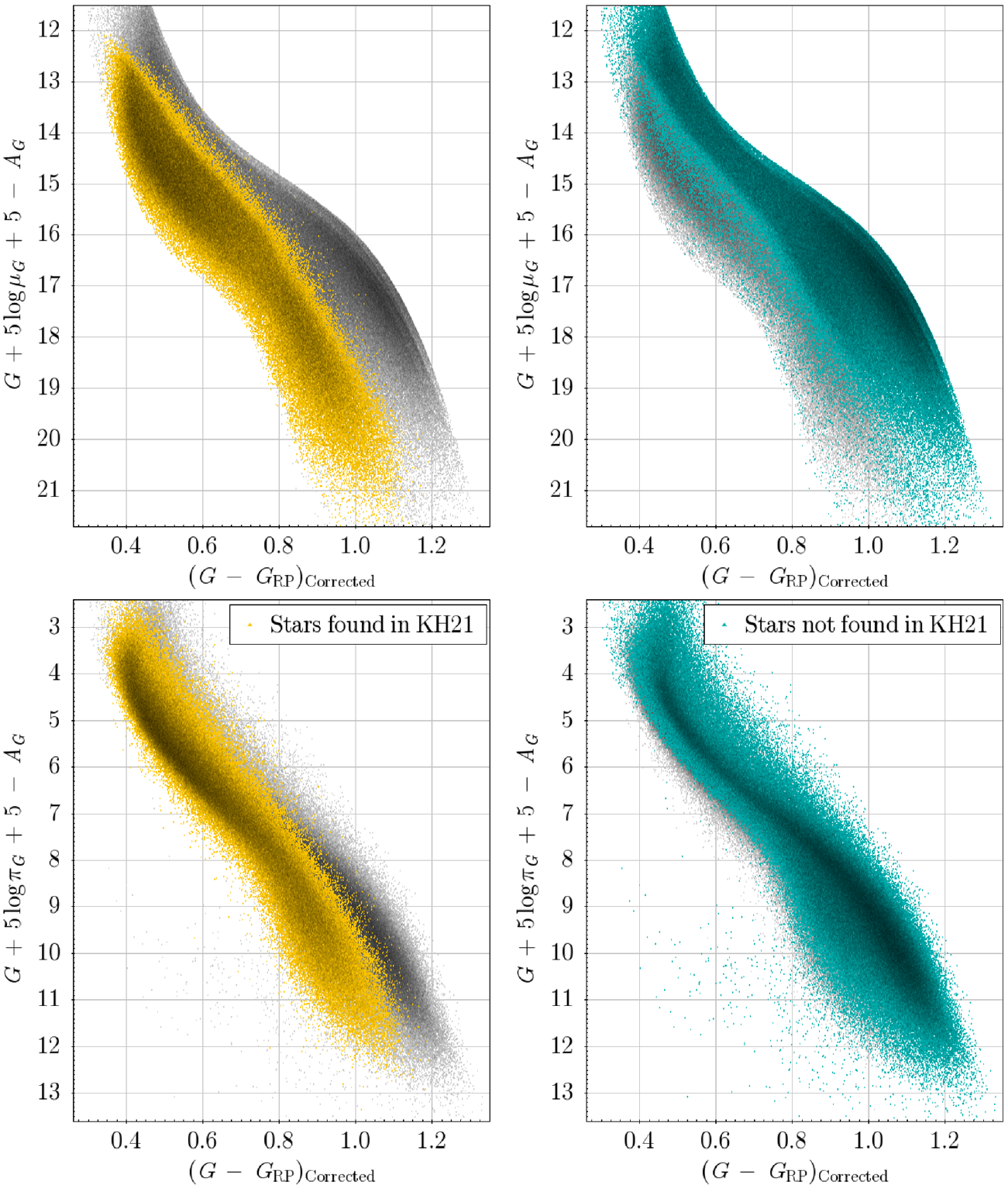}
\caption{Top panels: RPM diagrams of $205,655$ stars found in KH21 (left; yellow points) and $345,559$ stars not found in KH21 (right; blue points). Grey points are all stars in our catalogue. Stellar density anomaly around the edge of the distribution is caused by our selection cuts in different spatial areas. Stars not found in KH21 still shows the double sequences in the RPM diagrams, which indicates the mixture of the low-velocity halo stars in the thick disk population. Bottom panels: CMDs of stars found in KH21 (left) and stars not found in KH21 (right). Distribution of stars found in KH21 is clearly distinguished from the upper main sequence, where most of redder, metal-rich thick disk (compared to the halo population) stars dwell. 
\label{fig:fig19}}
\end{figure*}

However, KH21 ended up excluding stars on the red sequence in the CMD; our own more flexible selection has retained these redder objects, which happen to be less metal-poor. This difference is clear in the comparison between the RPM diagrams and CMDs of stars in each of the catalogues. We crossmatched our catalogue to KH21 and recovered $205,665$ stars in common between both catalogues. Figure~\ref{fig:fig19} shows the RPM (top panels) and colour-magnitude (bottom panels) diagrams of stars in our catalogue that are also listed in KH21 (left panels; yellow points) and stars in our catalogue not listed in KH21 (right panels; blue points). Grey points are all stars in our catalogue, shown for reference. The RPM diagram clearly shows the effects of the tight selection in KH21 which overlooks half the stars we selected. The consequences are quite apparent in the CMD, where much of the redder, less metal-poor stars (i.e., the red sequence) end up being excluded. 

The main issue is that the KH21 cut does not follow the colour-magnitude relationship of the local halo stars, and thus introduces a metallicity bias which is much stronger for low-mass stars; the result is that while KH21 includes a large fraction of the high-mass stars in our catalogue, most low-mass stars are not present in KH21, i.e., the local halo subset from KH21 is essentially ``top-heavy", and overlooks much of the lower-mass halo stars. It therefore appears that the RPM diagram selections are not entirely reliable for fully isolating halo late-type dwarfs from the thick disk population. While a strict cut may eliminate much of the disk stars, this may also end up eliminating many halo stars, especially among the less metal-poor stars of the halo. For example, among the lower-mass stars with $M_G > 7.5$ that were excluded in KH21, about $20$\% are stars that we identify to be metal poor objects ($\mathrm{[Fe/H]} < -1.2$) that we believe are most likely genuine halo members (see the bottom-right panel in Figure~\ref{fig:fig19}). Another way to see this is in the RPM distribution of stars not found in KH21 (top-right panel in Figure~\ref{fig:fig19}), which clearly includes low-mass stars from the metal-poor sequence. The conclusion is that limiting the selection to very nearby stars makes it possible to better tailor the selection in the RPM diagram and minimise metallicity biases.

\section{Summary and Conclusions}\label{sec:summary}
In this work, we have collected high-proper-motion stars with precise astrometric and photometric measurements in Gaia EDR3 in order to identify the local halo population out to 2 kpc. To identify the halo population, we use the RPM diagram, which is a tool to classify the local stellar populations only using given colours and proper motions. Since our selection is not dependent on distance, a reddening correction is essential to accurately identify the population. We adapt the 3D reddening map from \citet{green:19} and correct $G$ magnitudes and $\grp$ colours of stars in $b > -30^{\circ}$. The selection cut in the RPM diagram is set by the distribution of stars with high-tangential-velocities ($\vt > 200$ km s$^{-1}$). As the reddening map does not include the information on stars in southern hemisphere, the selection boundaries for the identification of halo stars from the RPM diagram are differently set up by six areas on the sky (see Figure~\ref{fig:fig1}).

Stellar metallicity is one of fundamental parameters that hold information on the formation and evolutionary history of a star and its surrounding environment. Thanks to large spectroscopic surveys, we are able to collect $20,047$ stars with spectroscopic metallicity measurements from the SDSS, LAMOST, GALAH, and \citet{hejazi:20}. We obtain metallicity estimates for our sample using the KNN regressor fitting, based upon the distribution of stars in the collective dataset as a training set. However, the estimates for low-mass dwarfs from the KNN regressor are overestimated due to the lack of data points in the training set. To overcome this problem, we built the photometric metallicity grid on the basis of the distribution of stars in the collective dataset. To check its credibility, we collected $322$ K+K and K+M wide binaries from \citet{hartman:20} and compared the metallicity estimates from the grid of primaries and secondaries. We applied a constant value ($0.5$ dex) to spectroscopic metallicities of late-type dwarfs in order to remove the systematic offset that we found between metallicity estimates of K+M pairs from the initial grid.

Our catalogue contains $551,214$ main-sequence stars in $d < 2$ kpc, including $321,879$ late-type dwarfs ($M_G > 7.5$). It is clear to see the dual main-sequence features in the CMD as same as the distribution of stars with high-tangential-velocity. Our grid line with $\mathrm{[Fe/H]_{grid}} = -1.2$ matches the gap between the upper and lower main sequence, however, this value is significantly lower than the result reported in \citet{gaia:20} who overplotted the \texttt{PARSEC} isochrone with $\mathrm{[Fe/H]} = -0.5$ and age of $11$ Gyr.

Stars in small areas near both Galactic poles and Galactic centre and anticentre present similar kinematic distribution in the projected Galactic velocity space as recent studies have reported. Although we do not present detailed chemodynamic analysis in this work, it is clear to find the two major local halo features in the kinematic distribution, which are the Gaia-Enceladus stream and the Helmi stream. We plan to present detailed chemodynamic analysis in future work.

Our catalogue will provide several possible targets for large spectroscopic surveys aiming to obtain observational data of the local halo population. The future Gaia data releases will provide 6D astrometric data only for stars with $G < 16$ as the Gaia radial velocity spectrometer \citep{cropper:18} adopts narrow passbands to minimise the background light. Therefore, this would be insufficient to obtain the full kinematics of more than $77$\% of stars in our catalogue if we only adopt the Gaia dataset to conduct the kinematic analysis. A better strategy would be to intentionally include them in the next large spectroscopic surveys, such as SDSS-V \citep{kollmeier:17}, 4MOST \citep{dejong:19}, WEAVE \citep{dalton:14}, or LSST \citep{ivezic:19}. As late-type dwarfs are known as the most common stellar population in the Universe, studying chemodynamics of halo late-type dwarfs will provide an interesting perspective on the evolutionary history of the Milky Way as the recent Gaia results have shown us.

\section*{Acknowledgements}
We are grateful to the anonymous referee for comments. We appreciate the Gaia team, SDSS APOGEE/SEGUE groups, and the LAMOST team for their efforts in producing the data sets we have used to calibrate our photometric metallicity grid.

This work has made use of data from the European Space Agency (ESA) mission
{\it Gaia} (\url{https://www.cosmos.esa.int/gaia}), processed by the {\it Gaia} Data Processing and Analysis Consortium (DPAC, \url{https://www.cosmos.esa.int/web/gaia/dpac/consortium}). Funding for the DPAC has been provided by national institutions, in particular the institutions participating in the {\it Gaia} Multilateral Agreement.

Funding for SDSS-III and SDSS-IV has been provided by the Alfred P. Sloan Foundation, the Participating Institutions, the National Science Foundation, and the U.S. Department of Energy Office of Science. SDSS-IV acknowledges support and resources from the centre for High Performance Computing at the University of Utah. The SDSS website is http://www.sdss.org.

Guoshoujing Telescope (the Large Sky Area Multi-Object Fiber Spectroscopic Telescope LAMOST) is a National Major Scientific Project built by the Chinese Academy of Sciences. Funding for the project has been provided by the National Development and Reform Commission. LAMOST is operated and managed by the National Astronomical Observatories, Chinese Academy of Sciences.

\section*{Data Availability}
The data underlying this article were accessed from the European Space Agency (ESA) mission {\it Gaia} (\url{https://www.cosmos.esa.int/gaia}), processed by the {\it Gaia} Data Processing and Analysis Consortium (DPAC,
\url{https://www.cosmos.esa.int/web/gaia/dpac/consortium}). The derived data generated in this research are available in the article and in its online supplementary material.

% Don't change these lines
\bsp	% typesetting comment
\label{lastpage}
\end{document}